\documentclass[prd, floatfix, nofootinbib,
preprintnumbers
,twocolumn,showpacs
]{revtex4}

\usepackage[utf8x]{inputenc}
\usepackage[T1]{fontenc}
\usepackage{lmodern}

\usepackage{amsmath}
\usepackage{amssymb}
\usepackage{bm}
\usepackage{epsfig}
\usepackage{graphicx}
\usepackage{color}
\usepackage{hyperref}
\usepackage{slashed}

\newcommand{\be}{\begin{equation}}
\newcommand{\ee}{\end{equation}}
\newcommand{\bdm}{\begin{displaymath}}
\newcommand{\edm}{\end{displaymath}}
\newcommand{\bea}{\begin{eqnarray}}
\newcommand{\eea}{\end{eqnarray}}
\newcommand{\nn}{\nonumber}

\newcommand{\flipped}{flipped $SU(5)$}
\newcommand{\TenFermion}{{10}_{M}}
\newcommand{\FiveFermion}{\overline{5}_{M}}
\newcommand{\OneFermion}{{1}_{M}}
\newcommand{\FiveHiggs}{5_H}
\newcommand{\FiveHiggsP}{5_H'}
\newcommand{\TenHiggs}{10_H}
\newcommand{\LHS}{W_\nu}
\newcommand{\PMNS}{V_{PMNS}}

\newcommand{\Yfive}{Y_{\overline{5}}}
\newcommand{\Yten}{Y_{10}}
\newcommand{\Yone}{Y_1}
\newcommand{\YtenP}{Y_{10}'}
\newcommand{\YfiveP}{Y_{\overline{5}}'}
\newcommand{\YoneP}{Y_1'}
\newcommand{\mz}[3]{\ensuremath{\left\{#1; #2 ; #3 \right\}}}
\newcommand{\mza}[1]{\ensuremath{\left\{#1\right\}}}
\newcommand{\mb}[1]{\ensuremath{\left[#1\right]}}

\DeclareMathOperator{\diag}{diag}

\DeclareMathOperator{\Tr}{Tr}
\DeclareMathOperator{\Br}{Br}
\DeclareMathOperator{\dilog}{Li_2}

\newcommand{\secref}[1]{Section~\ref{#1}}
\newcommand{\appref}[1]{Appendix~\ref{#1}}
\newcommand{\figref}[1]{\figurename~\ref{#1}}

\newcommand{\ipnp}{Institute of Particle and Nuclear Physics,
  Faculty of Mathematics and Physics,
  Charles University in Prague, V Hole\v{s}ovi\v{c}k\'ach 2,
  180 00 Praha 8, Czech Republic}

\graphicspath{{./figures/}}

\begin{document}

\title{Witten's loop in the minimal flipped $SU(5)$ unification revisited}
\preprint{}

\author{Dylan Harries}
\email{harries@ipnp.troja.mff.cuni.cz}
\affiliation{\ipnp}

\author{Michal Malinsk\'{y}}
\email{malinsky@ipnp.troja.mff.cuni.cz}
\affiliation{\ipnp}

\author{Martin Zdr\'{a}hal}
\email{zdrahal@ipnp.troja.mff.cuni.cz}
\affiliation{\ipnp}

\begin{abstract}
  In the simplest potentially realistic renormalizable variants of
  the flipped $SU(5)$ unified model the right-handed neutrino masses are
  conveniently generated by means of the Witten's two-loop mechanism.
  As a consequence, the compactness of the underlying scalar sector provides
  strong correlations between the low-energy flavor observables such as
  neutrino masses and mixing and the flavor structure of the fermionic currents
  governing the baryon and lepton number violating nucleon decays.
  In this study, the associated two-loop Feynman integrals are fully evaluated and, subsequently, are used to draw
quantitative conclusions about the central observables of interest such as the proton decay branching ratios and the absolute neutrino mass scale.\end{abstract}

\maketitle

\section{Introduction}
\label{sec:introduction}
Though not a genuine grand unified theory (GUT), the \flipped{} gauge
theory~\cite{DeRujula:1980qc,Barr:1981qv,Derendinger:1983aj} still attracts
significant attention~\cite{Ellis:2014xda,Cannoni:2015zmn,Sonmez:2016vem,Chen:2017rpn} due to several rather unique features it
exhibits.  In particular, one-stage symmetry breaking down to the Standard
Model (SM) can be achieved regardless of whether or not a TeV-scale
supersymmetry is assumed.  The corresponding Higgs sector can also be very
small, as it is sufficient to employ just a single $10$-dimensional
representation to accomplish the necessary symmetry breaking.  This is to
be compared to the $24$ of the Georgi-Glashow $SU(5)$~\cite{Georgi:1974sy}
and/or $45\oplus 16$ (or even $45\oplus 126$) of the minimal $SO(10)$ GUTs
(see, e.g., Refs.~\cite{Bertolini:2009es,Bertolini:2012im} and references
therein).

Flipped $SU(5)$ models also share several other nice features with their
truly unified cousins.  From the point of view of phenomenology, two such
features stand out as being particularly relevant due to their immediate
experimental consequences.  Firstly, as in the $SO(10)$ GUTs, 3 right-handed
(RH) neutrinos are enforced in the spectrum, allowing for the use of a
type-I seesaw mechanism to generate the light neutrino masses.  Additionally,
as in $SU(5)$ there is only one heavy gauge boson, which typically
yields somewhat stronger correlations between the flavor structure of the
baryon and lepton number violating (BLNV) currents and the low-energy flavor
observables, and hence one can often say quite a bit about, e.g., the
proton lifetime.

However, upon closer inspection one finds a certain level of tension between
the practical implications of these two points.  For example,
in order to implement the standard type-I seesaw with the RH neutrinos
at hand, a $50$-dimensional four-index scalar $50_S$ of $SU(5)$ is typically
added~\cite{Das:2005eb} together with a $3\times 3$ complex symmetric Yukawa
matrix $Y_{50}$ in order to generate the desired RH Majorana mass term via a
renormalizable coupling such as $Y_{50}^{IJ} 10_{FI}^{T} C^{-1} 10_{FJ} 50_{S}$.
Besides enlarging the scalar sector enormously (and, hence, disposing of the
uniquely small size of the ``minimal'' Higgs sector noted above as one of the
most attractive structural features of the framework), the extra scalar and
the associated Yukawa at play reduces the value of the low-energy neutrino
masses and the lepton mixing data as constraints for the proton lifetime
estimates as it essentially leaves the neutrino sector on its own.

Remarkably enough, this dichotomy may be overcome by
noticing~\cite{Leontaris:1991mq,Rodriguez:2013rma} that the RH neutrino masses
in \flipped{} models may be generated even without the unpleasant
extra $50_{S}$ at the two-loop level by means of a variant of the mechanism
first identified by Witten in the $SO(10)$ context~\cite{Witten:1979nr}.  The
two main features~\cite{Rodriguez:2013rma} of this scenario are, first, a
simple relation among the seesaw and the GUT scales where the former is, roughly
speaking, given by the latter times an extra two-loop suppression and, second,
a rigid correlation between the flavor structures of the neutrino and charged
sectors, which in most cases may be transformed into a set of strong constraints
for, e.g., the proton decay partial widths and branching ratios.

To this end, the Witten's-loop-equipped \flipped{} may even be viewed as
{\em the most economical renormalizable theory of the BLNV nucleon decays},
much simpler than, e.g., the potentially realistic variants of the $SO(10)$ and
even the $SU(5)$ GUTs.

From this perspective, it is interesting that in Ref.~\cite{Rodriguez:2013rma}
most of the basic features of this framework may have been identified even
without an explicit calculation of the graphs involved in Witten's mechanism.
In this work we intend to overcome this drawback by a careful inspection of
the Feynman graphs and their evaluation which, as we shall see, will clarify
several other points left unaddressed in the preceding studies.  In particular,
the calculation will make it very clear that the minimal potentially realistic
and renormalizable incarnation of the scheme under consideration is the variant
featuring a pair of $5$-dimensional scalars in the Higgs sector (besides a
single copy of the ``obligatory'' $10$-dimensional $\TenHiggs$ scalar). Second,
it will be shown that, in this framework, the light neutrino spectrum is always
forced to be on the heavy side (actually, within the reach of the KATRIN
experiment~\cite{Osipowicz:2001sq}), which, among other things, provides a
clear smoking gun signal of the scheme.

In \secref{sec:flipped-su5} we first provide a brief review of the \flipped{}
gauge theory context, identify the Feynman graphs underpinning the radiative RH
neutrino mass generation in the minimal and next-to-minimal models, and exploit the seesaw formula
in order to get strong constraints on their parameter space.
\secref{sec:calculation} is devoted to a detailed analysis of the relevant
two-loop graphs in the scenario with one copy of the 5-dimensional scalar in
the Higgs sector; this setting is simple enough to allow for a complete analytic
understanding of the results.  In \secref{sec:results} these findings are used
for the identification of the minimal potentially realistic model of this kind,
which is subsequently shown to be strongly constrained and potentially highly
predictive.  Most of the technical details of the lengthy calculations are
deferred to a set of appendices.

\section{Flipped $SU(5)$ \`a la Witten}
\label{sec:flipped-su5}
The defining feature of the \flipped{} unifications is the ``non-standard''
embedding of the SM hypercharge operator within its $SU(5)\otimes U(1)_{X}$
gauge symmetry algebra, namely
\be
Y=\frac{1}{5}(X-T_{24}) ,
\ee
where $T_{24}$ stands for the usual hypercharge-like generator of the
standard $SU(5)$ (normalized in such a way that the electric charge obeys
$Q=T_{L}^{3}+T_{24}$) and $X$ is the unique non-trivial anomaly-free generator
of the additional $U(1)$ normalized in such a way that it receives integer
values over the three basic irreps accommodating each generation of the
SM matter,
\be
\FiveFermion \equiv (\overline{5}, -3), \;\;
\TenFermion \equiv (10, +1), \;\;
\OneFermion \equiv (1, +5),
\ee
where the first number in brackets gives the $SU(5)$ representation and the
second the charge under $U(1)_X$.  Compared to the standard $SU(5)$ case,
the SM matter fields $u^c_L$ and $d^c_L$ are swapped with respect to their usual
assignments, i.e., the former is a member of $\FiveFermion$ while the latter
resides in $\TenFermion$.  Similarly, $e^c_L$ is found in the $SU(5)$ singlet
and the compulsory RH neutrino $\nu^c_L$ replaces it in the $10$-plet.

As for the gauge fields, the $(24,0)\oplus (1,0)$ adjoint of
$SU(5)\otimes U(1)_X$ in this context contains a multiplet $X_\mu$ transforming
under $SU(3)_C \otimes SU(2)_L \otimes U(1)_Y$ as $(3, \overline{2},
+\tfrac{1}{6})$, plus its hermitian conjugate, rather than the traditional
hypercharge-$\tfrac{5}{6}$ gauge bosons of the standard $SU(5)$.  The remaining
degrees of freedom account for the 12 SM gauge fields and one additional heavy
singlet.

The minimal Higgs sector sufficient for breaking the $SU(5)\otimes U(1)$
symmetry down to the SM and, subsequently, to the $SU(3)\otimes U(1)$ of QCD+QED
consists of $\TenHiggs = (10,+1)$\footnote{It may be worth
  pointing out here that, due to the non-zero $U(1)_{X}$ charge of
  $\TenHiggs$ inherent to the \flipped{} models, there is no way to build a
  non-renormalizable $d=5$ operator (presumably Planck-scale suppressed) that
  might, in the broken phase, affect the gauge-kinetic form and hence
  introduce significant theoretical uncertainties in the high-scale
  gauge-matching conditions and the determination of the GUT scale.
  As a result, one of the primary sources of irreducible uncertainties hindering
  the predictive power of the ``standard'' GUTs (such as the Georgi-Glashow
  $SU(5)$ or the non-minimal $SO(10)$ models with either $54$ or $210$
  breaking the unified symmetry) is absent from this class of models.}, in which
the SM singlet occupies the same position as the RH neutrino does in
$\TenFermion$, and $\FiveHiggs = (5,-2)$ containing the SM Higgs doublet.
The breakdown of $SU(5) \otimes U(1)_X$ to the SM gauge symmetry takes place
after the SM singlet present in $\TenHiggs$ develops a non-zero vacuum
expectation value (VEV), $V_G$, generating masses
\begin{equation} \label{eq:gauge-boson-masses}
  m_X^2 = \frac{g_5^2 V_G^2}{2}
\end{equation}
for the gauge bosons $X_\mu$, where $g_5$ is the $SU(5)$
gauge coupling.  The color triplet, $SU(2)_L$ singlet components of $\TenHiggs$
and $\FiveHiggs$ also mix at this stage to form a pair of massive color triplets
$\Delta_{1,2}$ transforming under the SM gauge symmetry as
$(3, 1, -\tfrac{1}{3})$, with masses $m_{\Delta_{1,2}}$.
Further details regarding the tree-level scalar spectrum in this minimal
\flipped{} model are given in \appref{app:spectrum}.

For the above embedding of the SM matter content and minimal set of Higgs
scalars, one can readily write the most general
renormalizable\footnote{Note that in non-renormalizable settings the benefits
  of the scheme may be lost as the Witten's loop contribution may be swamped
  by the effects of, e.g., the $d=5$ non-renormalizable operators of the
  $\TenFermion \TenFermion \TenHiggs \TenHiggs$ type.} Yukawa Lagrangian
(suppressing all flavor indices)
\be\label{Yukawa}
\mathcal{L} \ni \Yten \TenFermion \TenFermion \FiveHiggs
  + \Yfive \TenFermion \FiveFermion \FiveHiggs^{*}
  + \Yone \FiveFermion \OneFermion \FiveHiggs + h.c.\,,
\ee
with $\Yten$, $\Yfive$ and $\Yone$ denoting the relevant $3 \times 3$
complex Yukawa coupling matrices; note that the first of these, unlike the
latter two, is required to be symmetric in its flavor indices, i.e.,
$\Yten = \Yten^{T}$.  In the broken phase, the second term in
Eq.~(\ref{Yukawa}) yields a strong correlation among the Dirac neutrino mass
matrix $M_\nu^D$ and the up-type quark mass matrix $M_u$, namely,
\begin{equation} \label{eq:neutrino-dirac-equal-up}
  M_{\nu}^{D} = M_{u}^{T}
\end{equation}
at the GUT scale.  The flavor symmetric nature of
$\Yten$ also means that the down-type quark mass matrix satisfies
$M_d = M_d^T$, while the couplings in Eq.~(\ref{Yukawa}) say nothing specific
about the mass matrix $M_e$ of the charged leptons.  As we shall see, these
correlations will turn out to be central for the high degree of predictivity
of this framework\footnote{To this end, it is worth noting that these relations
  remain intact even in models with more than a single copy of $\FiveHiggs$
  in the scalar sector; as we shall see, this (especially the symmetry of
  $M_{d}$) will be crucial for the construction of the minimal potentially
  realistic scenario identified in \secref{subsec:two-five-plets}.}
entertained in the following sections.

\subsection{The RH neutrino masses and type-I seesaw}
So far, we have left aside any discussion of the physical light neutrino masses
in the current scenario.  Obviously, Eq.~(\ref{eq:neutrino-dirac-equal-up})
cannot be the whole story here and, thus, one has to employ a variant of the
seesaw mechanism in some way; since the type-II and/or type-III options cannot
be realized with the minimal scalar and fermionic sectors at hand one is
left with the type-I seesaw implemented through the Majorana mass term for
the RH neutrinos.

This may be most easily devised by employing a $50$-dimensional
scalar~\cite{Das:2005eb} that can couple to the $\TenFermion^{T} C^{-1}
\TenFermion$ fermionic bilinear; the VEV of a singlet therein then gives rise to the
desired mass term.  As was noted in \secref{sec:introduction}, however,
the associated single-purpose extra Yukawa matrix does not bring any
additional insight into the flavor structure of the model, and limits
the extent to which low-energy data can be used in constraining proton
decay observables.  Therefore we do not adopt this option here and, instead,
consider the effects emerging at the quantum level in the minimal model.

\subsubsection{The Witten's loop structure}
\begin{figure}[t]
  Topology 1: \parbox{5cm}{\includegraphics[width=4cm]{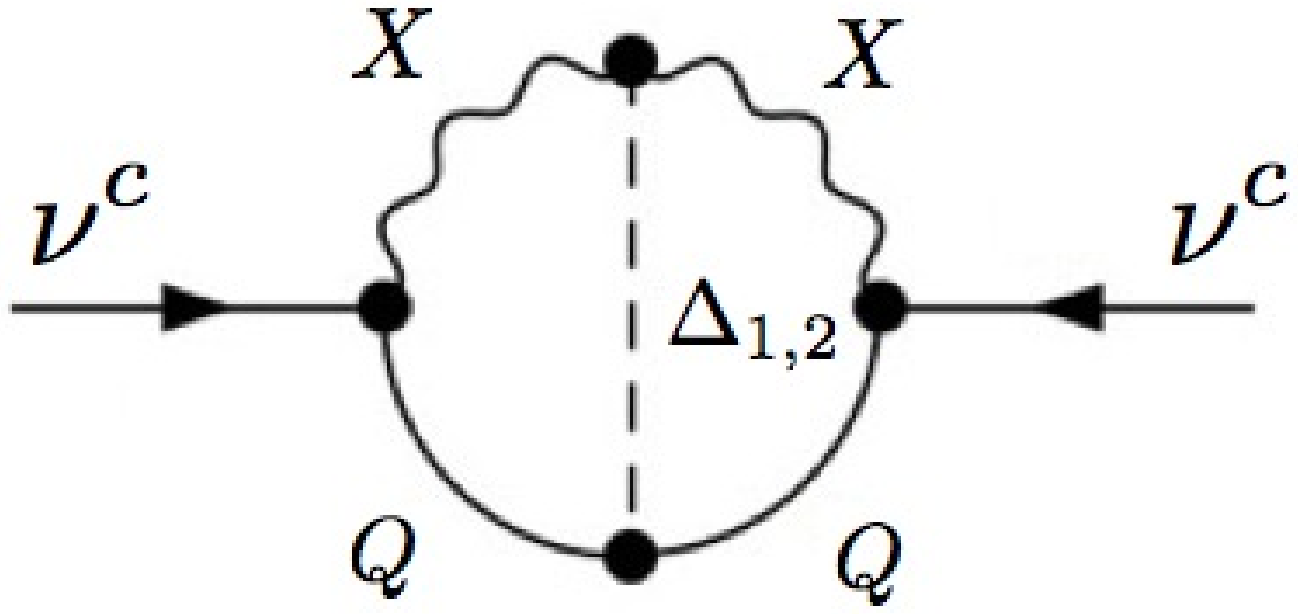}}
  \\
  Topology 2: \parbox{5cm}{\includegraphics[width=4cm]{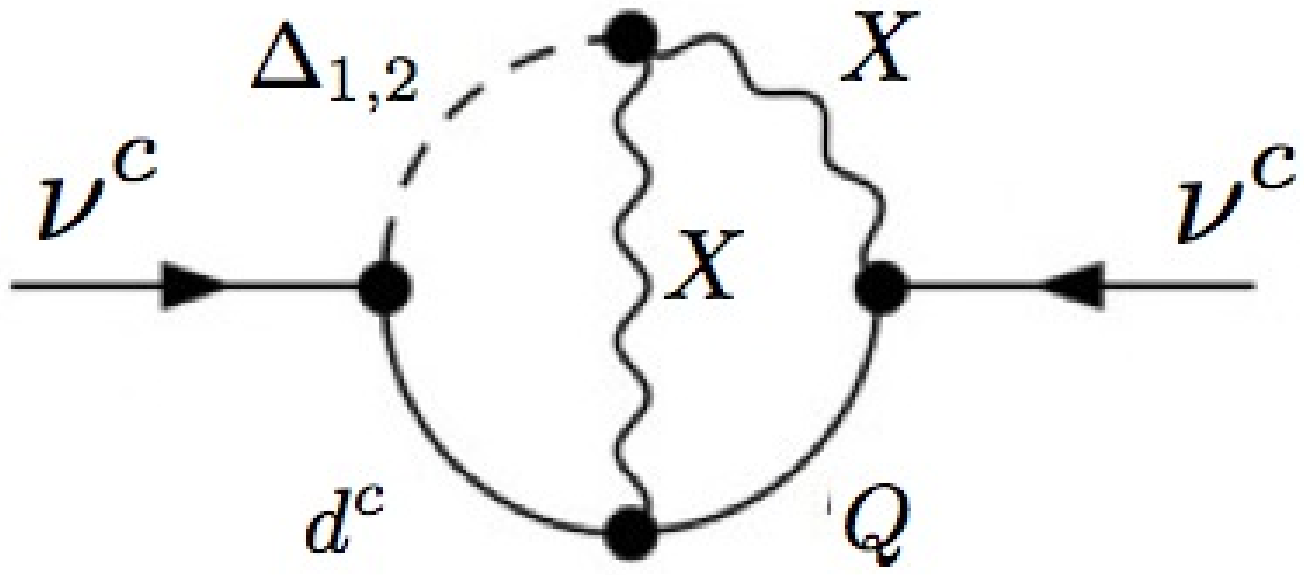}}
  \caption{\label{fig:graphs}The two non-equivalent topologies of the two-loop
    graphs contributing to the RH neutrino Majorana mass in the minimal
    \flipped{} model under consideration.  The vector field $X$ corresponds to
    the $({3},\overline{2},+\tfrac{1}{6})$ component of the adjoint while the
    pair of $\Delta$'s are the two mass eigenstates of the $(3,1,-\tfrac{1}{3})$
    colored scalars mixed from the relevant components of $\TenHiggs$ and
    $\FiveHiggs$, respectively.  
    }
\end{figure}

The simultaneous presence of the diquark-type of interactions, mediated
by the $X_{\mu}$ and $\Delta_{1,2}$ bosons, together with their
leptoquark counterparts (involving the same set of fields) in the model
implies that even $\Delta L=2$ Feynman diagrams corresponding to the Majorana
type of neutrino masses may be constructed at some higher order level.  This,
indeed, is the central point behind every radiative (Majorana) neutrino mass
generation mechanism; in the \flipped{} framework, it finds its incarnation in a
pair of two-loop topologies depicted in \figref{fig:graphs}, which can be viewed
as ``reduced'' versions of Witten's original $SO(10)$
graph(s)~\cite{Witten:1979nr}.

Note that in our analysis we shall work in the broken phase perturbation theory
with masses in the free Hamiltonian\footnote{Hence, we are avoiding the need to
  sum over an infinite tower of graphs (like the one drawn in Witten's original
  work~\cite{Witten:1979nr}) with increasing numbers of VEV insertions.  On the
  other hand, the explicit proportionality to the $\mu$ parameter governing the mixing
  between the $\TenHiggs$ and $\FiveHiggs$ multiplets (see
  \appref{app:spectrum}), which is obvious in the massless perturbation theory,
  becomes more involved in the massive case where $\mu$ emerges at the level of the relevant mixing matrix in the scalar sector, Eq.~(\ref{eq:su5-colour-triplet-mixing-elements}).} and in
the unitary gauge so that there are no Goldstone modes around.  This reduces
the number of relevant graphs considerably, albeit at the cost of making the
Feynman integrals somewhat more complicated compared to other cases.

Based on the graphs in \figref{fig:graphs} that remain in this approach, it is
immediately possible to make several comments on both the flavor structure and
overall scale of the generated Majorana mass matrix $M_\nu^M$.  The flavor
structure in particular plays a central role in what follows, and is governed
by the Yukawa couplings appearing in each of the contributing graphs.  In
each of the two topologies there is only a single Yukawa coupling
present, associated with the couplings of $\Delta_i$ to the fermions.  These
couplings involve only the $\FiveHiggs$ components of $\Delta_i$, since it is
only these components that couple to the fermions through the Yukawa
interactions in Eq.~(\ref{Yukawa}).  Moreover, since all of the fermions
appearing in the two graphs in \figref{fig:graphs} reside in $\TenFermion$,
the single relevant Yukawa coupling matrix is the symmetric $\Yten$.  Hence,
in the minimal model there is a tight correlation between the radiatively
generated RH neutrino Majorana mass matrix and the mass matrix of the
down-type quarks, making the scheme rather predictive.

The overall scale of $M_\nu^M$, on the other hand, depends on both the Yukawa
couplings in $\Yten$ as well as the gauge couplings and the sizes of the mass parameters
entering into each of the graphs.  One can initially estimate it to
be proportional to the dominant mass entry in the relevant graphs suppressed by
the appropriate two-loop factor and the combination of gauge (entering raised to
the fourth power) and Yukawa couplings.

Of the various mass parameters appearing in the evaluation of the graphs,
the fermionic masses $m_{f}$ should play no role in the integrals as the singlet
Majorana mass generation does not rely on the electroweak symmetry breaking.
Hence, in dealing with the Feynman integration we shall work in the chiral limit
with all SM fermions massless.  This, in principle, may lead to spurious IR
divergences in the form of, e.g., $\log(m_{f}/Q)$ arising in
individual partial fractions of the integrands, where $Q$ is the renormalization
scale, but as a whole $M_\nu^{M}$ should be stable in the $m_{f}\to 0$ limit.

Similarly, it is natural to expect that in the other extreme case corresponding
to one of the scalars $\Delta_{i}$ becoming significantly lighter with respect
to the $X_\mu$ boson masses (and, hence, bringing about another practically
massless propagator) $M_\nu^{M}$ should also remain regular; hence, the only
mass that can make it to the denominators in the final result is $m_{X}$.
This also suggests that, barring the couplings, each individual graph should
be governed by powers of the $m_{\Delta_{i}}/m_{X}$ ratio which, in turn, makes
it merely a function of a single\footnote{Assuming, implicitly, that the
  renormalization scale dependence eventually disappears as a consequence of
  the assumed UV-finiteness of the full result.} parameter.

\subsubsection{Seesaw as the key to the phenomenology}
Before coming to the evaluation of the graphs in \figref{fig:graphs} it is
important to stress that this is not all just an academic exercise; quite to
the contrary, the information obtained in \secref{sec:calculation} has a
profound impact on the phenomenology of the model.

The point is that, due to the seesaw formula, $M_\nu^{M}$ is correlated with the
physical light neutrino mass matrix $m_{LL}$ and the Dirac neutrino mass matrix
$M_\nu^{D}$ via
\be
M_\nu^{D} \left(m_{LL}\right)^{-1} (M_\nu^{D})^{T} = -M_\nu^M\, .
\ee
Using Eq.~(\ref{eq:neutrino-dirac-equal-up}), this can be
conveniently written as
\be\label{central}
\LHS \equiv D_{u} U_{\nu}^{\dagger} \left ( m_\nu^{\textrm{diag}} \right)^{-1}
U_{\nu}^{*}D_{u} = -M_\nu^M\,,
\ee
where $D_{u}$ is the diagonal form of the up-type quark mass matrix and
$U_{\nu}$ is the matrix diagonalizing $m_{LL}$, i.e., $m_{LL} = U_{\nu}^{T}
m_{\nu}^{\textrm{diag}} U_{\nu}$.  Note that in the derivation above we have
implicitly adopted the basis in which the up-type quark mass matrix is real
and diagonal, see Ref.~\cite{Rodriguez:2013rma} for further information.

Hence, up to an a priori unknown unitary matrix and the overall light neutrino
mass scale, parametrized for example by the mass of the heaviest of the light
neutrinos $m_{\nu}^{\textrm{max}}$, the matrix $\LHS$ defined in
Eq.~(\ref{central}) is completely determined by the low-energy quark masses
and neutrino oscillation data.  This is to be compared with $M_\nu^M$
appearing as the right-hand side of Eq.~(\ref{central}), which is set by
the heavy spectrum of the model (i.e., the masses of the heavy triplet
scalars and gauge bosons) and the gauge and Yukawa couplings, and
is therefore subject to other strong constraints.  In particular, $m_{X}$,
$m_{\Delta_{i}}$ and $g_5$ must be such that the unification pattern is
consistent with the low-energy data and compatible with the
non-observation of proton decay with at least $10^{34}$ years of lifetime~\cite{TheSuper-Kamiokande:2017tit}.

Hence, demanding consistency of Eq.~(\ref{central}) with the data one can
derive constraints on $m_{\nu}^{\textrm{max}}$ and, in particular, on $U_{\nu}$,
which is central to the BLNV phenomenology of the model.  Indeed, $U_{\nu}$
drives all the proton decay branching ratios into neutral mesons including
the ``golden channel'' $p\to \pi^0 e^{+}$ final state:
\begin{align}
  \nn
    \frac{\Gamma(p\to \pi^0 e_\alpha^+)}{\Gamma(p\to \pi^+\overline{\nu})} &=
    \frac{1}{2}|(V_{CKM})_{11}|^2|(\PMNS U_\nu)_{\alpha 1}|^2\,, \\
  \label{gamma4}
    \frac{\Gamma(p\to \eta e_\alpha^+)}{\Gamma(p\to \pi^+\overline{\nu})} &=
    \frac{C_2}{C_1}|(V_{CKM})_{11}|^2|(\PMNS U_\nu)_{\alpha 1}|^2\,, \\
  \nn
    \frac{\Gamma(p\to K^0 e_\alpha^+)}{\Gamma(p\to \pi^+\overline{\nu})} &=
    \frac{C_3}{C_1}|(V_{CKM})_{12}|^2|(\PMNS U_\nu)_{\alpha 1}|^2\,,
\end{align}
where the $C_i$'s are various low-energy factors calculable using chiral
Lagrangian techniques (see, e.g., Ref.~\cite{Nath:2006ut} and references
therein) and $V_{CKM}$ and $V_{PMNS}$ are the Cabibbo-Kobayashi-Maskawa
and the Pontecorvo-Maki-Nakagawa-Sakata mixing matrices, respectively.

In this sense,  the minimal \flipped{} unification equipped with the
  Witten's loop mechanism can be viewed as a particularly simple (if not
  the most minimal of all) theory of the absolute neutrino mass scale and,
  at the same time, the two-body BLNV nucleon decays.

\subsection{Consistency constraints and implications}
Let us now work out the aforementioned consistency constraints in more
detail and give some basic examples of their possible implications.
Firstly, it should be noted that there is a lower limit on the largest
entry of $W_\nu$ that depends on $m_\nu^{\textrm{max}}$ and the shape of
$U_\nu$.  Taking into account the typical $50\%$ reduction of the
running top quark Yukawa between $M_{Z}$ and the unification scale (at
around $10^{16}$ GeV) and taking, for example, $m_{\nu}^{\textrm{max}}=1$ eV
and $U_{\nu}=1$ one finds that the $(3,3)$ entry of $W_\nu$ is as large
as about
\be\label{LHStypicallowerlimit}
\left|(\LHS)_{33}\right|\sim 6.4\times 10^{12} \text{ GeV}.
\ee
The same magnitude, however, may not so easily be achieved for the $(3,3)$
entry of $M_\nu^M$ as required by Eq.~(\ref{central}) due to the generic
$10^{-3}$ geometrical suppression in the relevant two-loop graphs and a
possible further suppression associated with the Yukawa coupling $Y_{10}$;
the latter may be especially problematic in the minimal scenario~(\ref{Yukawa}) because
then $Y_{10}$ is fixed by the down-type quark masses and, thus, brings about
another suppression of some $10^{-2}$ to $(M_\nu^M)_{33}$.

However, this correlation
is loosened if there is more than a single copy of $\FiveHiggs$ in the
scalar sector.
As was already indicated in Ref.~\cite{Rodriguez:2013rma}, the additional $\YtenP$ associated to an extra  $\FiveHiggsP$ can conspire with the original $\Yten$ to do two things at once: they may partially cancel
in the down-type quark mass formula to account for the moderate suppression
of $M_{d}/M_{Z}$ yet their other combination governing $M_\nu^{M}$ (weighted by
the appropriate scalar mixings) may still remain large, thus avoiding the
problematic additional $10^{-2}$ suppression.  In what follows, we shall
model this situation by imposing a humble $|y|\lesssim 4\pi$ perturbativity
criterion on all the $\Yten$ and $\YtenP$ entries.

However, even in such a case the $\sim 10^{13}$ GeV lower limit on
the largest entry $(W_\nu)_{33}$, may still be problematic because,
for $U_{\nu}\neq 1$, it may be further enhanced by the admixture of the yet
larger $(2,2)$ and, in particular, the $(1,1)$ entry of
$\left(m_\nu^{\textrm{diag}}\right)^{-1}$; as a matter of fact the latter is not
constrained at all given that the lightest neutrino mass eigenstate may still be
extremely light. Thus, the lower bound on the magnitude of the largest
element of $W_\nu$ gets further boosted over the
na\"{\i}ve estimate of $10^{13}$ GeV whenever $U_{\nu}$ departs
significantly from unity, which in turn constrains all of the partial widths,
Eqs.~(\ref{gamma4}).

Hence, a thorough evaluation of the graphs in \figref{fig:graphs} will decide
several important questions, namely:
\begin{enumerate}
\item  Can the elements of $M_\nu^M$ ever be big enough to be consistent
  (at least in the most optimistic scenario with $U_{\nu}\sim 1$) with $W_\nu$,
  as required by Eq.~(\ref{central}), in the case of the single $\FiveHiggs$
  scenario with its typical extra $10^{-2}$ suppression at play?
\item If not, can the two-$\FiveHiggs$ scenario work? What would be then the
  corresponding lower limit for $m_{\nu}^{\textrm{max}}$ in this scenario?
\item In either case, what is the allowed domain for the entries of $U_{\nu}$
  and, thus, for the corresponding BLNV nucleon decay rates?
\end{enumerate}
This is what we turn our attention to in the remainder of this article.

\section{Witten's loop calculation}
\label{sec:calculation}
The leading contribution to the radiatively generated RH neutrino mass in the
current scheme may be computed by considering the  graphs in
\figref{fig:graphs} evaluated at zero external momentum, see
\appref{app:radiativemass}, with the relevant interaction terms given in
\appref{app:lagrangian}.  In the minimal renormalizable model containing only
a single $\TenHiggs$ and one or more $\FiveHiggs$ representations, no
one-loop contribution to the RH neutrino mass matrix can be generated, nor do
there exist any one-loop counterterm graphs.  The resulting expression for
the RH Majorana neutrino mass matrix in the case of a single $5_{H}$ multiplet reads
\begin{equation} \label{eq:rh-neutrino-mass-matrix}
  (M_\nu^M)^{IJ} = -\frac{3 g_5^4}{(4\pi)^4} V_G
  \sum_{i = 1}^2 (-8 \Yten^{IJ}) (U_\Delta)_{i1} (U_\Delta^*)_{i2} I_3(s_i) ,
\end{equation}
where the scalar mixing matrix elements $(U_\Delta)_{ij}$ are given in
\appref{app:spectrum}, and $I_3(s_i)$ is the sum of the
corresponding loop integrals evaluated at zero external incoming momentum,
\begin{equation} \label{eq:integral-sum-definition}
  I_3(s_i) = -(4 \pi)^4 ( \Sigma_1^P(0) + 2 \Sigma_2^P(0) ) ,
\end{equation}
regarded as a function of $s_i = m_{\Delta_i}^2 / m_X^2$. Recall that there is an overall extra factor of 2 included in Eq.~(\ref{eq:rh-neutrino-mass-matrix}) related to the permutation of the two external neutral field lines (for $I=J$) or to the symmetry of $Y_{10}$ (for $I\neq J$).  The integrals
$\Sigma_1^P(0)$ and $\Sigma_2^P(0)$, corresponding to topology 1 and 2
respectively, are given by
\begin{widetext}
\begin{align}
  i \Sigma^P_1(0) &= i \int \frac{d^4p}{(2\pi)^4} \int \frac{d^4q}{(2\pi)^4}
    \gamma_\rho \frac{1}{-\slashed{q}} \frac{1}{\slashed{p}} \gamma_\mu
    \frac{1}{(p+q)^2 - m_{\Delta_i}^2} \frac{-g^{\mu\nu}
    + \frac{1}{m_X^2} p^\mu p^\nu}{p^2 - m_X^2}\frac{-g_\nu^\rho
    + \frac{1}{m_X^2} q_\nu q^\rho}{q^2 - m_X^2} ,
    \label{eq:central-scalar-loop-integral} \\
  i \Sigma^P_2(0) &= i \int \frac{d^4p}{(2\pi)^4} \int \frac{d^4q}{(2\pi)^4}
    \frac{1}{-\slashed{q}} \gamma_\rho \frac{1}{\slashed{p}} \gamma_\mu
    \frac{1}{q^2 - m_{\Delta_i}^2} \frac{-g^{\mu\nu}
    + \frac{1}{m_X^2} p^\mu p^\nu}{p^2 - m_X^2} \frac{-g_\nu^\rho
    + \frac{1}{m_X^2}(p+q)_\nu(p+q)^\rho}{(p+q)^2 - m_X^2} .
    \label{eq:edge-scalar-loop-integral}
\end{align}
\end{widetext}
The integrals in
Eq.~(\ref{eq:central-scalar-loop-integral}) and
Eq.~(\ref{eq:edge-scalar-loop-integral}) are evaluated by reducing them
to expressions involving (variants of) the brackets by Veltman and van der Bij~\cite{vanderBij:1983bw},
which may be evaluated directly \cite{vanderBij:1983bw,
  Broadhurst:1987ei,Ford:1991hw,Ford:1992pn,Davydychev:1992mt,Sierra:2014rxa}.  The
details of this reduction, and the resulting analytic expressions for the
two-loop integrals, are given in \appref{app:brackets}.  In particular,
using the results given in Ref.~\cite{vanderBij:1983bw} and appropriate
generalizations thereof, it is found that the contributing brackets are
free of potential IR divergences in the limit of massless internal fermions,
such that the fermion masses may safely be allowed to vanish as in
Eq.~(\ref{eq:central-scalar-loop-integral})
and Eq.~(\ref{eq:edge-scalar-loop-integral}).  On the other hand, each graph
is individually UV divergent.  Setting $\epsilon = 2 - \frac{D}{2}$, where $D$
is the spacetime dimensionality, the divergences are found to be
\begin{equation} \label{eq:central-scalar-divergences}
  -(4\pi)^4 \Sigma_1^{P,\text{div}}(0) = \frac{3}{2 \epsilon}
  - \frac{m_{\Delta_i}^4}{2 m_X^4} \left ( \frac{1}{2\epsilon^2}
  + \frac{3}{2 \epsilon} - \frac{1}{\epsilon}
  \log \frac{m_{\Delta_i}^2}{Q^2} \right ) ,
\end{equation}
and
\begin{equation} \label{eq:edge-scalar-divergences}
  -(4 \pi)^4 \Sigma_2^{P,\text{div}}(0) = -\frac{3}{4\epsilon} +
  \frac{m_{\Delta_i}^4}{4 m_X^4} \left ( \frac{1}{2 \epsilon^2}
  + \frac{3}{2\epsilon} - \frac{1}{\epsilon} \log \frac{m_{\Delta_i}^2}{Q^2}
  \right ).
\end{equation}
It follows from Eq.~(\ref{eq:integral-sum-definition}) that the total
contribution $I_3(s_i)$ to the RH neutrino mass matrix is UV finite, as must
be the case here due to the absence of the necessary counterterms.

\section{Results}
\label{sec:results}
\begin{figure}[t]
  \includegraphics[width=0.8\columnwidth]{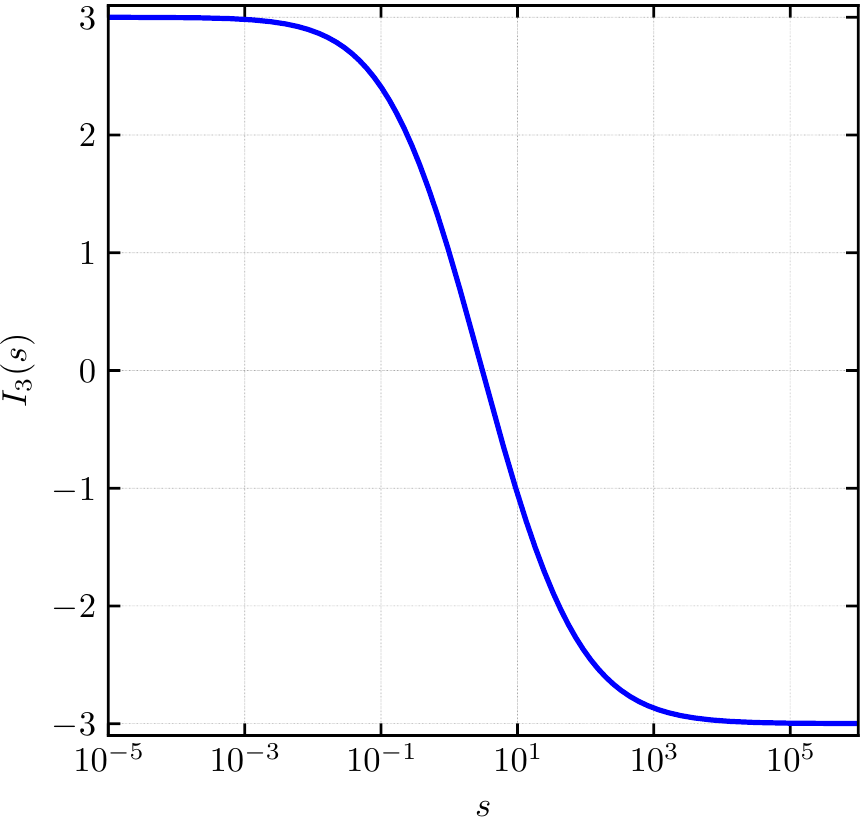}
  \caption{\label{fig:integral}Plot of the function $I_3(s)$ appearing in
  the RH neutrino mass matrix.}
\end{figure}

The behavior of the result for the purely kinematic piece of the RH neutrino
mass matrix, $I_3(s)$, is shown in \figref{fig:integral}.
Notably, the magnitude of $I_3(s)$ is bounded for all $s \geq 0$.  Indeed, from
the analytic result given in Eq.~(\ref{eq:I3intermsofs}), one has that for
$s \to 0$,
\begin{equation} \label{eq:integral-small-s-limit}
  I_3(s \to 0) = 3 + s \left ( 3 \log s + \pi^2 - \frac{15}{2} \right )
  + O(s^2 \log^2 s) ,
\end{equation}
while in the opposite limit with $s \to \infty$,
\begin{equation} \label{eq:integral-large-s-limit}
  I_3(s \to \infty) = -3 + O(s^{-1} \log^2 s) .
\end{equation}

\subsection{RH neutrino masses in the minimal model}
\label{subsec:one-five-plet}
\begin{figure}[t]
  \includegraphics[width=0.8\columnwidth]{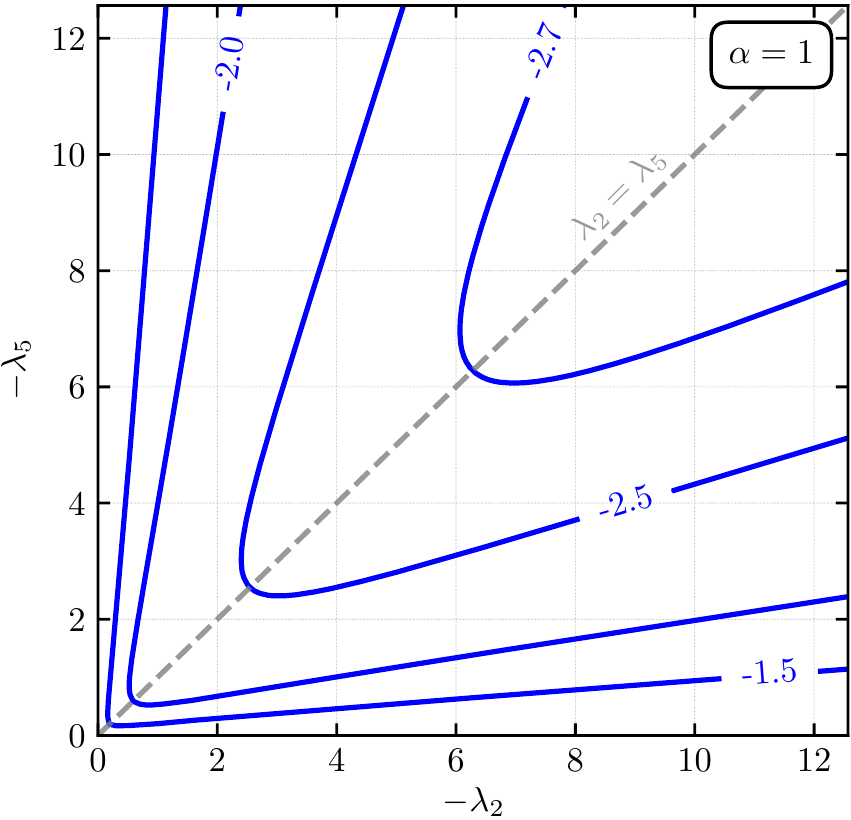}
  \caption{\label{fig:mass-matrix-sum-contours}Contour plot of $\tilde{I}$,
    as defined in Eq.~(\ref{eq:integral-sum}), in the
    $(\lambda_2, \lambda_5)$-plane, with $g_5 = 0.5$ and
    $\nu = \alpha \sqrt{\lambda_2 \lambda_5}$ for $\alpha = 1$, corresponding
    to the maximal value of $|\nu|$ consistent with a locally stable SM
    vacuum.}
\end{figure}

With $I_3(s)$ determined, we may proceed to evaluate the size of $M_{\nu}^{M}$
in Eq.~(\ref{eq:rh-neutrino-mass-matrix}).  Substituting in the explicit forms
of the mixing matrix elements in
Eq.~(\ref{eq:su5-colour-triplet-mixing-elements}) one obtains
\begin{equation} \label{eq:rh-neutrino-mass-matrix-2}
  M_\nu^M = -\frac{3 g_5^4}{(4\pi)^4} (-8 Y_{10}) V_G \tilde{I} ,
\end{equation}
where
\begin{equation} \label{eq:integral-sum}
  \tilde{I} = \sum_{i = 1}^2
  \frac{2\nu^* \left ( 2\lambda_2 + {g_5^2 s_i} \right )}
  {4|\nu|^2 + \left ( 2\lambda_2 + {g_5^2 s_i} \right )^2} I_3(s_i) ,
\end{equation}
and $\nu = \mu / V_G$.  We note that $\tilde{I} \to 0$ as $\mu \to 0$,
reflecting the fact that the graphs rely on the $\TenHiggs - \FiveHiggs$
mixing.  It is also clear from Eq.~(\ref{eq:integral-sum}) that,
since $I_3(s)$ is bounded, $\tilde{I}$ cannot be made arbitrarily large
to compensate for the suppression factors noted in \secref{sec:flipped-su5}.
To develop some sense of the allowed size of $\tilde{I}$, it is useful to
substitute for $s_{i}$ from Eq.~(\ref{eq:su5-colour-triplet-scalar-masses2})
and inspect $\tilde{I}$ as a function of $\nu$, $\lambda_2$, $\lambda_5$, and
$g_5$, neglecting all terms that are of the order of $v^2 / V_G^2$, where
$v$ is the electroweak VEV, see Eq.~(\ref{eq:vev-definition}).  Requiring
that the tree-level vacuum be locally stable implies \cite{Rodriguez:2013rma}
$\lambda_{2,5} < 0$ and
\begin{equation} \label{eq:vacuum-stability-constraint}
  |\nu| \leq \sqrt{\lambda_2 \lambda_5} .
\end{equation}
When this bound is saturated, i.e., when $|\nu| = \sqrt{\lambda_2 \lambda_5}$,
the mass $m_{\Delta_1}$ vanishes for all values of $\lambda_2$, $\lambda_5$
while $m_{\Delta_2}^2 =-( \lambda_2 + \lambda_5)V_{G}^{2}$.
The resulting value of $\tilde{I}$ for this special case is shown
in the $(\lambda_2, \lambda_5)$ plane in \figref{fig:mass-matrix-sum-contours}.
In particular, it should be noted that the value of $\tilde{I}$ is unchanged
under the interchange $\lambda_2 \leftrightarrow \lambda_5$, as can be easily
verified from Eqs.~(\ref{eq:integral-sum}) and
(\ref{eq:su5-colour-triplet-scalar-masses2}), and $|\tilde{I}| \leq 3$ for all
values of $\lambda_2$ and $\lambda_5$.  The maximal value of $|\tilde{I}|$ is
achieved for $\lambda_2 = \lambda_5$, with $|\tilde{I}| \to 3$ as
$\lambda_2 = \lambda_5 \to -\infty$.

\begin{figure}[th]
  \includegraphics[width=0.8\columnwidth]{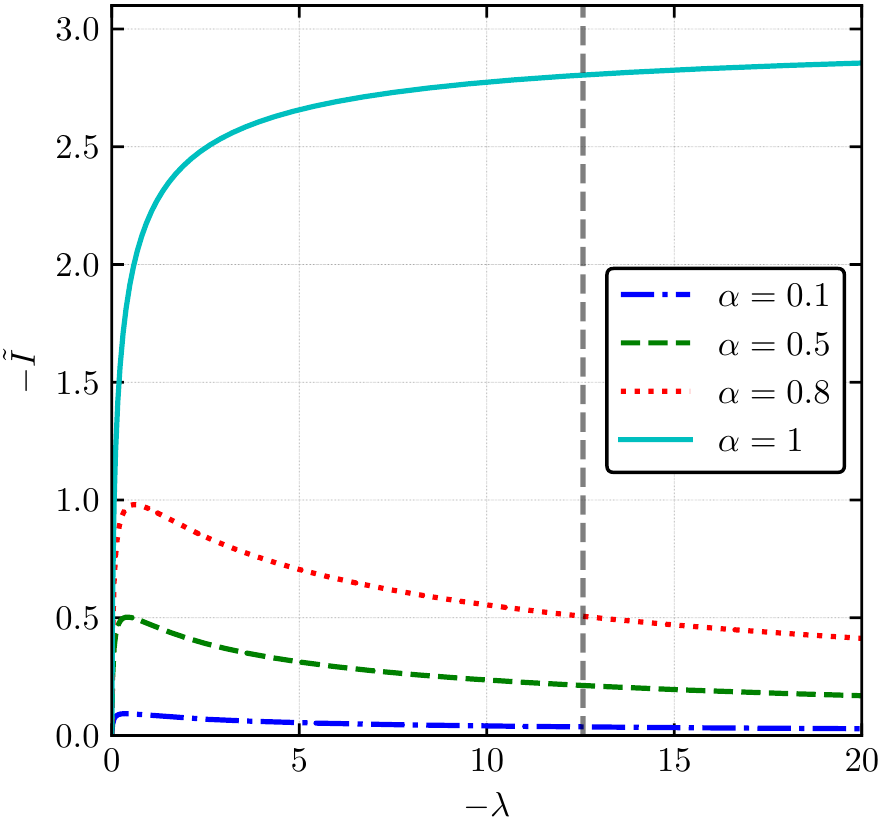}
  \caption{\label{fig:mass-matrix-sum}Plot of the range of variation of
    $\tilde{I}$ as a function of $\lambda_2 = \lambda_5 = \lambda$, with
    $g_5 = 0.5$ and $\mu = \alpha \sqrt{\lambda_2 \lambda_5}
    V_G$, for $\alpha \in [0, 1]$.  The dashed vertical line denotes the
    na\"{\i}ve perturbativity limit $|\lambda_i| \leq 4\pi$.}
\end{figure}

Qualitatively different behavior results in the more general case that $\nu$
does not saturate the bound given in Eq.~(\ref{eq:vacuum-stability-constraint}).
This is demonstrated in \figref{fig:mass-matrix-sum}, in which the value of
$\tilde{I}$ is plotted as a function of $\lambda_2 = \lambda_5 = \lambda$
with
\begin{equation} \label{eq:nu-constraint-fraction}
  \nu = \alpha \sqrt{\lambda_2 \lambda_5} , \quad \alpha \in [0, 1] ,
\end{equation}
for several values of $\alpha$.
Although $\tilde{I}$ remains invariant under $\lambda_2 \leftrightarrow
\lambda_5$, with the maximum value of $|\tilde{I}|$ still occurring for
$\lambda_2 = \lambda_5$, for values of $|\alpha| < 1$, $|\tilde{I}|$ now
tends to zero for large values of the scalar couplings $\lambda_2$, $\lambda_5$.
This is due to the fact that, for $|\alpha| \neq 1$, both $s_1$, $s_2$ grow
with increasing $|\lambda|$ such that $I_3(s_1), I_3(s_2) \to -3$, while the
coefficients of each in Eq.~(\ref{eq:integral-sum}) are equal in magnitude but
of opposite sign, resulting in the two terms cancelling.  Physically, this
corresponds to the expected dynamical decoupling of the heavy scalar states in
the $m_{\Delta_{1,2}} \to \infty$ limit.  For $\alpha = 1$, at least
one color triplet scalar is massless at tree-level for all values of $\lambda_2$
and $\lambda_5$.  Consequently, this state never decouples and $\tilde{I}$
therefore does not vanish.  Technically, this arises because $I_3(s_1) = 3$
while $I_3(s_2) \to -3$, with the two contributions still entering 
$\tilde{I}$ with coefficients of equal magnitude but opposite sign.

However, even in the most optimistic case with $|\tilde{I}| \to 3$, the above
results make it clear that there is little hope for a viable prediction of the
light neutrino spectrum in the minimal scenario under consideration.
For acceptable values of $m_X \sim 10^{17}$ GeV, and taking $g_5 \approx 0.5$,
the elements of $M_\nu^M$ are found to be $\lesssim 10^{12}$ GeV after taking
into account the $\sim 10^{-2}$ suppression associated with presence of $\Yten$.
This is to be compared with the (optimistic) lower bound of $\sim 10^{13}$
GeV for the elements of the left-hand side of Eq.~(\ref{central}).
Evidently, in the case when only a single $\FiveHiggs$ is present in the
spectrum the answer to whether Eq.~(\ref{central}) can be satisfied is negative.
In fact, in this minimal model the problem is exacerbated by the fact that
$\Yten \propto M_d$, which implies a far too hierarchical pattern of light
neutrino masses irrespective of their absolute size, as was previously
noted in Ref.~\cite{Rodriguez:2013rma}.  Thus we are immediately led to
consider the remaining questions raised in \secref{sec:flipped-su5}
concerning the viability of the model with an additional $\FiveHiggs$
representation instead.

\subsection{Minimal potentially realistic model}
\label{subsec:two-five-plets}
As noted above, the addition of a second $\FiveHiggs$ multiplet in principle
allows both the $\Yten$ suppression and the overly hierarchical flavor structure
to be avoided.  At the same time, the overall
predictive power of the theory is not significantly harmed by this addition; in
particular, doing so does not spoil the key Yukawa relations
used in obtaining Eq.~(\ref{central}).  With a
second $\FiveHiggsP$ multiplet, the Yukawa sector of the model reads
\begin{align}
  \mathcal{L} &\ni \Yten \TenFermion \TenFermion \FiveHiggs
    + \YtenP \TenFermion \TenFermion \FiveHiggsP \nonumber \\
  & \quad {} + \Yfive \TenFermion \FiveFermion \FiveHiggs^*
    + \YfiveP \TenFermion \FiveFermion \FiveHiggsP^* \nonumber \\
  & \quad {} + \Yone \FiveFermion \OneFermion \FiveHiggs
    + \YoneP \FiveFermion \OneFermion \FiveHiggsP + h.c.,
    \label{eq:two-five-yukawa-sector}
\end{align}
where $\YtenP$ is of course also flavor symmetric.  In this scenario, the
Dirac neutrino mass matrix still remains tightly correlated with the up-type
quark masses, with the GUT scale relation
\begin{equation} \label{eq:two-five-dirac-neutrino-mass-matrix}
  M_\nu^D = M_u^T \propto \Yfive v + \YfiveP v'
\end{equation}
holding at tree-level, where $v'$ is the VEV associated with the electrically
neutral component of $\FiveHiggsP$, see \appref{subapp:two-five-spectrum}.  By
contrast, the analogous relationship between the down-type quark masses and
the generated RH neutrino Majorana masses, $M_d, M_\nu^M \propto \Yten$, is
no longer preserved.  While $M_d \propto \Yten v + \YtenP v'$, the appropriate
generalization of Eq.~(\ref{eq:rh-neutrino-mass-matrix}) reads
\begin{equation} \label{eq:two-five-rh-neutrino-mass-matrix}
  (M_\nu^M)^{IJ} = -\frac{3 g_5^4}{(4 \pi)^4} V_G \sum_{i = 1}^3 \sum_{j = 2}^3
  (-8 Y_j^{IJ}) (U_\Delta)_{i1} (U_\Delta^*)_{ij} I_3(s_i) ,
\end{equation}
where $Y_j = \Yten$ when $j = 2$ and $Y_j = \YtenP$ when $j = 3$,
with $U_\Delta$ now a $3 \times 3$ mixing matrix as defined
in Eq.~(\ref{eq:two-5-scalar-triplet-mass-eigenstates}).  Thus, in general,
$M_d$ and $M_\nu^M$ are determined by different linear combinations of
the Yukawa couplings $\Yten$ and $\YtenP$.  In turn, this means that the
generic suppression of $M_\nu^M$ by a factor $\propto M_d$ may be
avoided in the two-$\FiveHiggs$ scenario.
On the other hand, it is still the case that the elements
of $M_\nu^M$ are bounded from above, at least so long as it is required that
all couplings remain perturbative.

\subsubsection{Phenomenology of the minimal potentially realistic model}
As the ignorance of yet higher-order effects makes any such perturbativity
constraints somewhat arbitrary in general, in what follows we shall give two
examples of the $M_{\nu}^{M}$ estimates corresponding to two different choices
of the upper limits on the effective (running) SM down-quark
Yukawa couplings.  These, according to  Eq.~(\ref{eq:LYukawas}), obey
$Y_{d}\equiv 8 \Yten$ and $Y_{d}'\equiv 8 \YtenP$ at the matching scale. The
two cases to be considered are i) $|Y_{d}|,|Y_{d}'| \lesssim 1$
and ii) $|Y_{d}|,|Y_{d}'| \lesssim 4\pi$.  For the former case (motivated by
the SM value of the top Yukawa coupling) one has the following upper limit on
$M^{M}_{\nu}$ calculated from Eq.~(\ref{eq:two-five-rh-neutrino-mass-matrix})
\begin{equation}
  \text{case i)}\quad
  \left|M^{M}_{\nu}\right|\lesssim 6.4 \times 10^{12} \left(
  \frac{m_{X}}{10^{17} \rm GeV} \right)\, {\rm GeV}, \label{limit1}
\end{equation}
while for the latter one obtains
\begin{equation}
  \text{case ii)}\quad
  \left|M^{M}_{\nu}\right|\lesssim 8.0 \times 10^{13} \left(
  \frac{m_{X}}{10^{17} \rm GeV}\right)\, {\rm GeV} . \label{limit2}
\end{equation}

Note that in both cases we have used the (numerical) upper limit
\begin{equation}
  \left|\sum_{i = 1}^3 \sum_{j = 2}^3
  (U_\Delta)_{i1} (U_\Delta^*)_{ij} I_3(s_i)\right|\leq 3
\end{equation}
which is completely analogous to the limit discussed in
\secref{subsec:one-five-plet} for the single-$\FiveHiggs$ case.

Remarkably, for the typical \flipped{} value of  $m_{X}=10^{17}$ GeV (see,
e.g., Ref.~\cite{Rodriguez:2013rma}) the case i) limit, Eq.~(\ref{limit1}),
is just on the borderline of compatibility with the optimistic lower
limit in Eq.~(\ref{LHStypicallowerlimit}) on $|W_{\nu}|$, while
the latter case ii) in principle admits lower\footnote{These, however, may
  not be that simple to get within potentially realistic unification chains,
  see Appendix C of Ref.~\cite{Rodriguez:2013rma}.} values of $m_{X}$.

This, in turn, implies that there is generally not much room for any
significant admixture of the second neutrino (inverse) mass within the
element $(W_\nu)_{33}$, hence, the only allowed $U_{\nu}$'s in
Eq.~(\ref{central}) are those for which $(U_\nu)_{13}$ and $(U_\nu)_{23}$ are
small.

To this end, the model clearly calls for a dedicated numerical analysis
including a detailed calculation of the heavy spectrum that conforms to, among
other things, the requirement of a significant spread of the scalar triplets in
order to maximize $|\tilde{I}|$.  This, however, is beyond the
scope of the current study and will be elaborated on elsewhere.

At this point, let us just illustrate the typical situation by evaluating
the most significant proton-decay two-body branching ratios (neglecting the
kinematically suppressed vector-meson channels for simplicity) in the
$(U_{\nu})_{13}=(U_{\nu})_{23}=0$ limit with the 1-2 mixing angle $\theta_{12}$
therein chosen in such a way that $\Gamma(p\to \pi^{0}\mu^+)$ is maximized
(see Ref.~\cite{Rodriguez:2013rma} for further details):
\begin{equation}
  \begin{aligned}
    \Br(p\to \pi^{+}\overline{\nu})&\approx 80.0\% \, , \\
    \Br(p\to \pi^{0}e^{+}) &\approx 14.2\% \, , \\
    \Br(p\to \pi^{0}\mu^{+}) &\approx 5.5\% \, , \\
    \Br(p\to K^{0}e^{+}) &\approx 0.1\% \, .
  \end{aligned}
\end{equation}
Needless to say, for non-extremal values of  $\theta_{12}$ these branching
ratios may vary; in particular, $\Br(p\to \pi^{0}e^{+}) /
\Br(p\to \pi^{0}\mu^{+})$ should increase.

Finally, let us say a few words about the lower limits on the mass of the heaviest SM neutrino in the two cases~(\ref{limit1}) and~(\ref{limit2}). As for the former, one obtains\footnote{Given the structure of the seesaw formula in the current context~(\ref{central}) together with the tight constraints on the structure of the $U_{\nu}$ matrix we generally assume the hierarchy of the light neutrino mass eigenstates to be normal.} 
\be
m_{3}\gtrsim \left(
  \frac{10^{17} \rm GeV}{m_{X}} \right)\,{\rm eV}
\ee
while for the latter one has
\be
m_{3}\gtrsim 0.08 \left(
\frac{10^{17} \rm GeV}{m_{X}} \right)\,{\rm eV}
\ee
which, actually, turns out to be independent on the specific form of the $U_{\nu}$ matrix as long as the 1-3 and 2-3 mixings therein are small (see the discussion above). 
With this at hand, any specific experimental upper limit on the absolute neutrino mass scale may be readily translated into a lower limit on $m_{X}$ and, subsequently, the proton lifetime.

\section{Conclusions and outlook}
\label{sec:conclusion}
The two-loop radiative RH neutrino mass generation mechanism originally
identified by Witten in 1980s in the $SO(10)$ context finds a beautiful
incarnation in the class of renormalizable \flipped{} unified theories
where, among other effects, it avoids the need for the $50$-dimensional scalar
representation.  This, in turn, renders the simplest potentially realistic
scenarios perhaps the most minimal (partially) unified gauge theories on the
market, with strong implications for some of the key beyond-Standard-Model
observables such as the absolute neutrino mass scale and proton decay.

In this work we have focused on a thorough evaluation of the relevant Feynman
graphs in these scenarios paying particular attention to their analytic
properties and the absolute size of the effect which turns out to be the key
to the consistency of the scenario as a whole.  It has been shown that there is
no way to be consistent with the data with only one $5$-dimensional scalar
multiplet at play and, hence, the minimal potentially realistic setup must
include two such irreps in the scalar sector (along with the $10$-dimensional
tensor).

As it turns out, such a minimal \flipped{} model is subject to strong
constraints on its allowed parameter space that lead to rather stringent
limits on the absolute light neutrino mass scale as well as the BLNV
two-body nucleon decays.  A thorough numerical analysis of the corresponding
correlations is deferred to a future study.

\section*{Acknowledgments}
The authors acknowledge financial support from the Grant Agency of the
Czech Republic (GA\v{C}R) under the contract no. 17-04902S. We would like to thank Helena Kole\v{s}ov\'{a}, Ji\v{r}\'{i} Novotn\'{y}, Catarina Sim\~{o}es and Diego Aristizabal Sierra for illuminating discussions.

\appendix

\section{The interaction Lagrangian}
\label{app:lagrangian}
The radiative generation of the RH neutrino masses involves only a small
subset of the interactions associated with the full \flipped{}
Lagrangian.  Working in the $SU(5) \otimes U(1)_X$ broken phase, we extract
the required interactions from the kinetic terms and general Yukawa Lagrangian,
Eq.~(\ref{Yukawa}), making use of \texttt{FeynRules}
\cite{Christensen:2008py,Alloul:2013bka} and \texttt{FeynArts}
\cite{Kublbeck:1990xc,Hahn:2000kx} to verify that all terms and contributing
diagrams are accounted for.  As discussed in \secref{sec:flipped-su5},
when the model contains only a single $\FiveHiggs$ representation
the relevant diagrams are found to be those in \figref{fig:graphs}, arising
from the interaction Lagrangian
\begin{align}
  \mathcal{L}_{\text{int}} &\ni \frac{g_5^2}{2}
    \epsilon_{ijk}  \epsilon^{\beta\alpha} V_G X^{\mu i}_{\alpha}
    X_{\mu\beta}^{j} \overline{D}^{\dagger k} + \frac{g_5}{\sqrt{2}}
    \epsilon_{ijk} X_{\mu \alpha}^{i} \overline{d_{L_I}^c}^{j} \gamma^\mu
    Q_{L_I}^{k \alpha} \nonumber \\
  & \quad {} + \frac{g_5}{\sqrt{2}} \epsilon^{\beta \alpha}
    X_{\mu \alpha}^{i} \left ( \overline{Q_{L_I}} \right )_{i\beta} \gamma^\mu
    \nu_{L_I}^c - 8 \Yten^{IJ} d_{L_Ii}^{cT} C^{-1} \nu_{L_J}^c T^{i}
    \nonumber \\
  & \quad {} - 4 \Yten^{IJ}  \epsilon_{ijk}  \epsilon_{\alpha\beta}
    (Q_{L_I}^{i\beta})^T C^{-1} Q_{L_J}^{j\alpha} T^{k}
    + h.c. \label{eq:interaction-lagrangian}
\end{align}
where $i$, $j$, $k$ and $\alpha$, $\beta$ denote the $SU(3)_{C}$ and
$SU(2)_{L}$ indices, respectively, and $\epsilon_{ijk}$ and
$\epsilon_{\alpha\beta}$ are the relevant fully antisymmetric tensors with
$\epsilon_{123} = -\epsilon^{12} = 1$.  In this expression, $\overline{D}$
denotes the $(\bar{{3}}, {1}, +\frac{1}{3})$ components of the scalar
$\TenHiggs$, $T$ the $({3},{1},-\frac{1}{3})$ components of $\FiveHiggs$,
$Q_{L_I}$ the quark doublet $({3},{2}, +\frac{1}{6}) \in \TenFermion$,
$d_{L_I}^c$ the down-type quark singlet $(\bar{{3}}, {1},
+\frac{1}{3}) \in \TenFermion$, and $\nu_{L_I}^c$ the $({1}, {1}, 0)$
components of $\TenFermion$.  The charged vector bosons $X_\mu$ associated
with the breaking of $SU(5) \otimes U(1)_X$ have SM quantum numbers
$({3}, \bar{{2}}, +\frac{1}{6})$.  Following the breakdown of the
$SU(5) \otimes U(1)_X$ symmetry due to the non-zero VEV $V_G$, the scalar
states $\overline{D}$ and $T$ mix to form the $SU(3)_C \otimes SU(2)_L
\otimes U(1)_Y$ eigenstates $\Delta_{1,2}$, as described in
\appref{app:spectrum}.

Let us note that in deriving the central formula
Eq.~(\ref{eq:rh-neutrino-mass-matrix}), especially the overall factor of $3$
therein, the color and isospin factors in Eq.~(\ref{eq:interaction-lagrangian})
play a crucial role.  It is also worth noting that the exact cancellation of
the UV divergences discussed in \secref{sec:calculation}, which relies on the
extra factor of $2$ in Eq.~(\ref{eq:integral-sum-definition}), emerges from
the difference of the overall numerical factors in the last two terms in
Eq.~(\ref{eq:interaction-lagrangian}).

After including an additional $\FiveHiggsP$ to arrive at the minimal realistic
model discussed in \secref{subsec:two-five-plets}, the interaction Lagrangian
remains rather similar.  The addition of Yukawa couplings involving
$\FiveHiggsP$ leads to the set of interaction terms (with color indices
suppressed for simplicity)
\begin{align}
  \mathcal{L}^{\text{TH5M}}_{\text{int}} &=
  \mathcal{L}_{\text{int}} - \Big [ 8 (\YtenP)^{IJ} d_{L_I}^{cT} C^{-1}
    \nu_{L_J}^c T'
    \label{eq:two-5-interaction-lagrangian}
    \\
    & \quad {} + 4 (\YtenP)^{IJ} \epsilon_{\alpha\beta} (Q_{L_I}^\beta)^T C^{-1}
    Q_{L_J}^\alpha T'  + h.c. \Big ] , \nn
\end{align}
where $T'$ denotes the additional $({3}, {1}, -\frac{1}{3})$
multiplet contained in $\FiveHiggsP$, which mixes with the states
$\overline{D}$ and $T$ to yield a set of $SU(3)_C \otimes SU(2)_L
\otimes U(1)_Y$ eigenstates $\Delta_{1,2,3}$.

For the sake of completeness and matching to the SM Yukawa couplings we also
present the terms involving the doublet Higgs interactions here:
\begin{align}
  -\mathcal{L}_{\text{int}} &\ni 8 \Yten^{IJ} \epsilon_{\gamma \delta} H^\delta
    d^{cT}_{L_I} C^{-1} Q^\gamma_{L_J}
    + \Yfive^{IJ} H_\gamma^\dagger u^{cT}_{L_J} C^{-1} Q^\gamma_{L_I}
    \nonumber \\
  & \quad {} + \Yfive^{IJ} H^\dagger_\gamma \nu^{cT}_{L_I} C^{-1}
    \ell^\gamma_{L_J}
    + \Yone^{IJ} \epsilon_{\gamma \delta} H^\gamma e^{cT}_{L_J} C^{-1}
    \ell^\delta_{L_I} \nonumber \\
  & \quad {} + h.c., \label{eq:LYukawas}
\end{align}
where the SM Higgs doublet $H$ consists of the components of $\FiveHiggs$
transforming under the SM gauge group as $({1}, {2}, -\frac{1}{2})$,
$u^c_{L_I}$ and $\ell_{L_I}$ are the components of $\FiveFermion$
transforming as $(\overline{{3}}, {1},
-\frac{2}{3})$ and $({1}, {2}, -\frac{1}{2})$ respectively, and
$e^c_{L_I}$ denotes the single component of $\OneFermion$, transforming as
$({1}, {1}, +1)$.

\section{Triplet scalar spectrum and mixing}
\label{app:spectrum}
\subsection{Model with a single $\FiveHiggs$ representation}
\label{subapp:one-five-spectrum}
The tree-level scalar potential in the model with a single $\FiveHiggs$ may be
written
\begin{align}
  V &= \frac{1}{2} m_{10}^2 \Tr ( \TenHiggs^\dagger \TenHiggs )
    + m_5^2 \FiveHiggs^\dagger \FiveHiggs \nonumber \\
  & \quad {} + \frac{1}{8} \left (
    \mu \epsilon_{ijklm} \TenHiggs^{ij} \TenHiggs^{kl} \FiveHiggs^m + h.c.
    \right ) \nonumber \\
  & \quad {} + \frac{1}{4} \lambda_1 \left [
    \Tr ( \TenHiggs^\dagger \TenHiggs ) \right ]^2
    + \frac{1}{4} \lambda_2 \Tr ( \TenHiggs^\dagger \TenHiggs
    \TenHiggs^\dagger \TenHiggs )
    \nonumber \\
  & \quad {} + \lambda_3 ( \FiveHiggs^\dagger \FiveHiggs )^2
    + \frac{1}{2} \lambda_4 \Tr ( \TenHiggs^\dagger \TenHiggs )
    ( \FiveHiggs^\dagger \FiveHiggs )
    \nonumber \\
  & \quad {} + \lambda_5 \FiveHiggs^\dagger \TenHiggs \TenHiggs^\dagger
    \FiveHiggs . \label{eq:minimal-scalar-potential}
\end{align}
The scalar basis is chosen such that the spontaneous breaking of
$SU(5) \otimes U(1)_X$ and the subsequent electroweak symmetry breaking
takes place via the non-zero VEVs
\begin{equation} \label{eq:vev-definition}
  \langle \TenHiggs \rangle^{45} = -\langle \TenHiggs \rangle^{54} = V_G , \quad
  \langle \FiveHiggs \rangle^4 = v .
\end{equation}
Requiring that this corresponds to a stationary point of the scalar
potential yields the conditions
\begin{align}
  V_G \left [ m_{10}^2 + V_G^2 ( 2 \lambda_1 + \lambda_2 )
    + v^2 ( \lambda_4 + \lambda_5 ) \right ] &= 0 ,
  \label{eq:ewsb-cond-1} \\
  v \left [ m_5^2 + 2 \lambda_3 v^2 + V_G^2 ( \lambda_4 + \lambda_5 )
    \right ] &= 0 ,
  \label{eq:ewsb-cond-2}
\end{align}
which permit the parameters $m_5^2$, $m_{10}^2$ to be eliminated in
favor of the VEVs.

After the breakdown of $SU(5) \otimes U(1)_X$ to $SU(3)_C \otimes SU(2)_L
\otimes U(1)_Y$, the charged vector bosons $X_\mu$ associated with the
broken generators acquire masses $m_X$ given by
Eq.~(\ref{eq:gauge-boson-masses}).  The scalar states $T$ and
$\overline{D}$ of relevance to the generation of the RH neutrino
masses mix, with the mass matrix (in the basis $(\overline{D}^\dagger, T)$)
\begin{equation} \label{eq:su5-colour-triplet-mass-matrix}
  M_{\Delta}^2 = \begin{pmatrix}
    -\lambda_2 V_G^2 & \mu V_G \\
    \mu^* V_G & m_5^2 + \lambda_4 V_G^2
  \end{pmatrix} ,
\end{equation}
where Eq.~(\ref{eq:ewsb-cond-1}) with $v = 0$ has been used to eliminate
$m_{10}^2$.  This is diagonalized by a unitary matrix $U_\Delta$ according to
\begin{equation*}
  U_\Delta M_\Delta^2 U_\Delta^\dagger = \begin{pmatrix} m_{\Delta_1}^2 & 0 \\
    0 & m_{\Delta_2}^2
  \end{pmatrix} ,
\end{equation*}
with
\begin{align} \label{eq:su5-colour-triplet-scalar-masses}
  m_{\Delta_{1,2}}^2 & = \frac{1}{2} \Bigl\{ m_5^2 + \left ( \lambda_4
  - \lambda_2 \right ) V_G^2 \\ \nonumber
  & \mp \sqrt{\left [ m_5^2 + \left ( \lambda_2
      + \lambda_4 \right ) V_G^2 \right ]^2 + 4 |\mu|^2 V_G^2} \Bigr\},
\end{align}
which, in the electroweak vacuum, simplifies into
\begin{equation} \label{eq:su5-colour-triplet-scalar-masses2}
  m_{\Delta_{1,2}}^2  = \frac{V_G^2}{2 } \Bigl\{-\left ( \lambda_2
  + \lambda_5 \right )   \mp \sqrt{\left ( \lambda_2
      - \lambda_5 \right )^{2}  + \frac{4|\mu|^2}{V_G^2 }} \Bigr\}.
\end{equation}
The elements of the mixing matrix $U_\Delta$ read
\begin{equation}
  \begin{aligned}
    (U_\Delta)_{11} &= \frac{\mu^* V_G}{\sqrt{|\mu|^2 V_G^2
        + \left ( m_{\Delta_1}^2 + \lambda_2 V_G^2 \right )^2}} , \\
    (U_\Delta)_{12} & = \frac{m_{\Delta_1}^2
      + \lambda_2 V_G^2}{\sqrt{|\mu|^2 V_G^2
        + \left ( m_{\Delta_1}^2 + \lambda_2 V_G^2 \right )^2}}
    , \\
    (U_\Delta)_{21} &= \frac{\mu^* V_G}{\sqrt{|\mu|^2 V_G^2
        + \left ( m_{\Delta_2}^2 + \lambda_2 V_G^2 \right )^2}} ,\\
    (U_\Delta)_{22} &= \frac{m_{\Delta_2}^2
      + \lambda_2 V_G^2}{\sqrt{|\mu|^2 V_G^2
        + \left ( m_{\Delta_2}^2 + \lambda_2 V_G^2 \right )^2}} .
  \end{aligned} \label{eq:su5-colour-triplet-mixing-elements}
\end{equation}

\subsection{Model with two $\FiveHiggs$ representations}
\label{subapp:two-five-spectrum}
In the minimal realistic model with two $\FiveHiggs$ representations, we take
the tree-level scalar potential to be given by
\begin{align}
  V &= \frac{1}{2} m_{10}^2 \Tr ( \TenHiggs^\dagger \TenHiggs )
      + m_{5}^2 \FiveHiggs^\dagger \FiveHiggs
      + m_{5'}^2 \FiveHiggsP^\dagger \FiveHiggsP \nonumber \\
    & \quad {} + \frac{1}{4} \lambda_1 \left [ \Tr ( \TenHiggs^\dagger \TenHiggs
      ) \right ]^2
      + \frac{1}{4} \lambda_2 \Tr ( \TenHiggs^\dagger \TenHiggs
      \TenHiggs^\dagger \TenHiggs ) \nonumber \\
    & \quad {} + \lambda_3 ( \FiveHiggs^\dagger \FiveHiggs )^2
      + \tilde{\lambda}_3 ( \FiveHiggsP^\dagger \FiveHiggsP )^2
      + \lambda_6 ( \FiveHiggs^\dagger \FiveHiggsP )
      ( \FiveHiggsP^\dagger \FiveHiggs ) \nonumber \\
    & \quad {} + \tilde{\lambda}_6 ( \FiveHiggs^\dagger \FiveHiggs )
      ( \FiveHiggsP^\dagger \FiveHiggsP )
      + \frac{1}{2} \lambda_4 \FiveHiggs^\dagger \FiveHiggs
      \Tr ( \TenHiggs^\dagger \TenHiggs ) \nonumber \\
    & \quad {} + \frac{1}{2} \tilde{\lambda}_4 \FiveHiggsP^\dagger \FiveHiggsP
      \Tr ( \TenHiggs^\dagger \TenHiggs )
      + \lambda_5 \FiveHiggs^\dagger \TenHiggs \TenHiggs^\dagger
      \FiveHiggs \nonumber \\
    & \quad {} + \tilde{\lambda}_5 \FiveHiggsP^\dagger \TenHiggs
      \TenHiggs^\dagger \FiveHiggsP + \Big [ m_{12}^2 \FiveHiggs \FiveHiggsP
      \nonumber \\
    & \quad {} + \frac{\mu}{8} \epsilon_{ijklm} \TenHiggs^{ij} \TenHiggs^{kl}
      \FiveHiggs^m
      + \frac{\mu'}{8} \epsilon_{ijklm} \TenHiggs^{ij} \TenHiggs^{kl}
      \FiveHiggsP^m \nonumber \\
    & \quad {} + \eta_1 ( \FiveHiggs^{\dagger} \FiveHiggs )
      ( \FiveHiggs^{\dagger} \FiveHiggsP )
      + \eta_2 ( \FiveHiggs^{\dagger} \FiveHiggsP )^2 \nonumber \\
    & \quad {} + \eta_3 ( \FiveHiggs^{\dagger} \FiveHiggsP )
      ( \FiveHiggsP^{\dagger} \FiveHiggsP )
      + \frac{1}{2} \lambda_7 \FiveHiggs^\dagger \FiveHiggsP
      \Tr ( \TenHiggs^\dagger \TenHiggs ) \nonumber \\
    & \quad {} + \lambda_8 \FiveHiggs^\dagger \TenHiggs \TenHiggs^\dagger
      \FiveHiggsP + h.c. \Big ] . \label{eq:two-5-scalar-potential}
\end{align}
The field basis is again chosen such that the fields $\TenHiggs$ and
$\FiveHiggs$ acquire non-zero VEVs given by Eq.~(\ref{eq:vev-definition}),
while
\begin{equation} \label{eq:second-5-vev-definition}
  \langle \FiveHiggsP \rangle^4 = v' .
\end{equation}
The corresponding conditions that must hold for this to be a stationary
point of the potential are
\begin{equation} \label{eq:two-5-ewsb-conditions}
  f_i = 0 , \quad i = 1, 2, 3 ,
\end{equation}
where
\begin{align}
  f_1 &= v_1 m_{5}^2 + v_2 m_{12}^2 + 3 v_1^2 v_2 \eta_1
    + v_2^3 \eta_3 \nonumber \\
  & \quad {} + v_2 V_G^2 ( \lambda_7 + \lambda_8 )
    + 2 v_1^3 \lambda_3
    + v_1 V_G^2 ( \lambda_4 + \lambda_5 ) \nonumber \\
  & \quad {} + v_1 v_2^2 ( \lambda_6
  + \tilde{\lambda}_6 + 2 \eta_2 ) , \label{eq:two-5-ewsb-cond-1} \\
  f_2 &= v_2 m_{5'}^2 + v_1 m_{12}^2 + v_1^3 \eta_1
    + 3 v_1 v_2^2 \eta_3 \nonumber \\
  & \quad {} + v_1 V_G^2 ( \lambda_7 + \lambda_8 )
    + 2 v_2^3 \tilde{\lambda}_3
    + v_2 V_G^2 ( \tilde{\lambda}_4 + \tilde{\lambda}_5 )
    \nonumber \\
  & \quad {} + v_1^2 v_2 ( \lambda_6 + \tilde{\lambda}_6 + 2 \eta_2 ) ,
    \label{eq:two-5-ewsb-cond-2} \\
  f_3 &= V_G m_{10}^2 + V_G^3 ( 2 \lambda_1 + \lambda_2 )
  + v_1^2 V_G ( \lambda_4 + \lambda_5 ) \nonumber \\
  & \quad {} + v_2^2 V_G ( \tilde{\lambda}_4
  + \tilde{\lambda}_5 ) + 2 v_1 v_2 V_G ( \lambda_7
  + \lambda_8 ) . \label{eq:two-5-ewsb-cond-3}
\end{align}
In deriving the above, and in all expressions below, we restrict our
attention to the case where all couplings are real.

In the $SU(3)_C \otimes SU(2)_L \otimes U(1)_Y$ symmetric phase, i.e.,
for $V_G \neq 0$, $v = v' = 0$, the set of scalar color triplets that
mix is extended to include the color triplet $T'$ associated with $\FiveHiggsP$.
The $3 \times 3$ mass matrix, in the basis $(\overline{D}^\dagger, T, T')$,
reads
\begin{equation} \label{eq:two-5-su5-colour-triplet-mass-matrix}
  M_\Delta^2 = \begin{pmatrix}
      -\lambda_2 V_G^2 & \mu V_G & \mu' V_G \\
      \mu V_G & m_{5}^2 + \lambda_4 V_G^2 & m_{12}^2 + \lambda_7 V_G^2 \\
      \mu' V_G & m_{12}^2 + \lambda_7 V_G^2 &
      m_{5'}^2 + \tilde{\lambda}_4 V_G^2
  \end{pmatrix} ,
\end{equation}
where Eq.~(\ref{eq:two-5-ewsb-cond-3}) with $v = v' = 0$ has been used
to eliminate the dependence on $m_{10}^2$.  The resulting mass eigenstates
$(\Delta_1, \Delta_2, \Delta_3)$ are obtained through the rotation
\begin{equation} \label{eq:two-5-scalar-triplet-mass-eigenstates}
  \begin{pmatrix}
    \Delta_1 \\ \Delta_2 \\ \Delta_3
  \end{pmatrix} =
  U_\Delta \begin{pmatrix}
    \overline{D}^\dagger \\ T \\ T'
  \end{pmatrix} ,
\end{equation}
where the unitary matrix $U_\Delta$ diagonalizes $M_\Delta^2$
according to
\begin{equation}
  U_\Delta M_\Delta^2 U_\Delta^\dagger =
  \diag ( m_{\Delta_1}^2, m_{\Delta_2}^2, m_{\Delta_3}^2 ) .
\end{equation}

\section{Radiative fermion mass generation}
\label{app:radiativemass}
In general, the physical mass of a single spin-$1/2$ fermion is obtained as the
value of $m$ for which
\begin{equation} \label{eq:general-fermion-pole-mass}
  ( \slashed{k} + m ) \Gamma^{(2)}(k) = 0
  \quad \forall k \text{ such that } k^2 = m^2 ,
\end{equation}
where $\Gamma^{(2)}(k)$ is the renormalized two-point 1PI Green's function,
\begin{equation} \label{eq:two-point-1PI-greens-function}
  \Gamma^{(2)}(k) = Z(k) \slashed{k} - \Sigma(0) .
\end{equation}
In this expression, $Z(k)$ corresponds to the wavefunction renormalization and
$\Sigma(0)$ is the zero incoming momentum contribution to the appropriate
sum of Feynman diagrams.  Taken together,
Eq.~(\ref{eq:general-fermion-pole-mass}) and
Eq.~(\ref{eq:two-point-1PI-greens-function}) imply that
\begin{equation} \label{eq:pole-mass-condition}
  m Z(m^2) = \Sigma(0) ,
\end{equation}
which generally amounts to a transcendental equation to be solved for the
physical mass $m$.  An expression for $m$ may be obtained perturbatively by
writing $Z(m^2) = 1 + \Delta Z(m^2)$, $\Sigma(0) = m_0 + \Delta m_0$, where
the first and second term in each expression correspond to the tree-level
and loop corrections to each quantity, respectively.  One finds the result
\begin{equation} \label{eq:perturbative-physical-mass}
  m = m_0 + [ \Delta m_0 - m_0 \Delta Z(m_0^2) ] + \ldots ,
\end{equation}
where we show only the leading part of the higher-order contribution.
Therefore, in the general case with $m_0 \neq 0$, a calculation of the leading
higher-order contribution to the physical mass would require the evaluation of
the loop corrections to both $\Sigma(0)$ and $Z(k^2)$.

However, for the case studied in this article in which the RH neutrinos are
massless at tree-level, Eq.~(\ref{eq:perturbative-physical-mass}) reads simply
$m = \Delta m_0 = \Sigma(0)$ at leading order.

\section{Evaluation of the two loop Feynman integrals}
\label{app:brackets}

\subsection{Veltman-Van der Bij brackets}
Remarkably enough, there is an entire industry concerning the evaluation
methods for the zero-external-momentum two-point 1PI graphs, see, e.g.,
Ref.~\cite{vanderBij:1983bw} or Ref.~\cite{Sierra:2014rxa} and references
therein.

The principal object in these methods are the so-called Veltman-Van der Bij
brackets.  As the original paper uses an Euclidean metric and a different
choice of dimensional regularization parameter $\epsilon$, we give here all
of the relevant expressions in our particular convention, i.e., in Minkowski
metric $g = \diag(1,-1,-1,-1)$ and with the number of spacetime dimensions
equal to $D=4-2\epsilon$.

We introduce the brackets in the following way
\begin{widetext}
\begin{align}
  \mz{M_{11},M_{12},\dots}{M_{21},\dots}{M_{31},\dots} &=
    \int \frac{d^4p}{(2\pi)^4} \int \frac{d^4q}{(2\pi)^4}
    \frac{1}{(p^2 - M_{11}^2) (p^2 - M_{12}^2) \dots}
    \frac{1}{(q^2 - M_{21}^2) \dots}
    \frac{1}{[(p + q)^2 - M_{31}^2] \dots} \, , \\
  \mza{M_{11}, M_{12}, \dots} &=
    \int \frac{d^4p}{(2\pi)^4} \frac{1}{(p^2 - M_{11}^2) (p^2 - M_{12}^2) \dots}
    \, , \\
  \mz{M_{11}, \dots}{M_{21}, \dots}{M_{31}, \dots} \mb{A(p,q)} &=
    \int \frac{d^4p}{(2\pi)^4} \int\frac{d^4q}{(2\pi)^4}
    \frac{1}{(p^2 - M_{11}^2) \dots}
    \frac{1}{(q^2 - M_{21}^2)\dots}
    \frac{1}{[(p+q)^2 - M_{31}^2] \dots}A(p,q) \, .
\end{align}
\end{widetext}
With the last expression we have introduced a shorthand notation that
simplifies the form of this appendix.

Note that the brackets are invariant under the exchange of positions of the
individual groups of components, which can be obtained by the change of
variables $(p\leftrightarrow q)$ and $(p+q\to p, -q \to q)$.

By a partial cancellation of fractions we can derive various reduction
formulae of the type
\begin{equation}
\begin{split}
&\mz{M_A,m_a}{M_B,m_b}{M_C}\mb{p^2}\\
&=\mz{M_A}{M_B,m_b}{M_C}+m_a^2\mz{M_A,m_a}{M_B,m_b}{M_C}.
\end{split}
\end{equation}
A similar trick using $p^2-M_B^2-(p^2-M_A^2)=M_A^2-M_B^2$ can be used for a
simplification of brackets of the type\footnote{Note that this simplification
  relates together the Passarino-Veltman integrals $A_0$ and $B_0$,
  \begin{equation}
    B_0(0,0,M_A^2) =\mza{M_A,0} = \frac{1}{M_A^2}\mza{M_A}
    =\frac{1}{M_A^2}A_0(M_A^2).
  \end{equation}
}
\begin{equation}\label{eq:difference}
  \mz{M_A,M_B}{\alpha}{\beta}=\frac{1}{M_A^2-M_B^2}\left(
  \mz{M_A}{\alpha}{\beta}-\mz{M_B}{\alpha}{\beta}\right).
\end{equation}
It is also possible to show that
\begin{equation}
  \begin{split}
    &\mz{M_A}{M_B}{M_C}\mb{(p+q)^2}\\
    &\qquad\qquad=\mza{M_A}\mza{M_B}+M_C^2\mz{M_A}{M_B}{M_C}.
  \end{split}
\end{equation}

Using all of these methods we can express the relevant two-loop integrals in
terms of simple brackets $\mz{M_A}{M_B}{M_C}$.  It is of use to rewrite them
further into double brackets
\begin{equation}
  \mz{2M_A}{M_B}{M_C}\equiv \mz{M_A,M_A}{M_B}{M_C},
\end{equation}
which are dimensionless (cf. Ref.~\cite{vanderBij:1983bw}).
The operation transcribing simple brackets into double brackets is
't~Hooft's $p$-operation~\cite{tHooft:1972tcz}.  In our notation it reads
\begin{equation}\label{D-3}
  \begin{split}
    \mz{M_A}{M_B}{M_C}&= \frac{1}{D-3} \Big ( M_A^2\mz{2M_A}{M_B}{M_C}\\
    & \quad {} + M_B^2\mz{2M_B}{M_C}{M_A} \\
    & \quad {} + M_C^2\mz{2M_C}{M_A}{M_B} \Big).
  \end{split}
\end{equation}

\subsection{Topology 1}
Topology 1 of \figref{fig:graphs} leads to the kinematic form (i.e.,
neglecting the specific form of the vertices) of the integral given in
Eq.~(\ref{eq:central-scalar-loop-integral}).  By using $D$-dimensional gamma
matrix gymnastics, it can be simplified into
\begin{equation}
  \begin{split}
    \Sigma^P_1(0) &= -\mz{m_X,0}{m_X,0}{m_{\Delta_i}} \Big [
       (D-4) \slashed{q} \slashed{p} + 4 p \cdot q \\
      & \quad {} - \frac{p^2 + q^2}{m_X^2} \slashed{p} \slashed{q}
      + \frac{p^2 q^2}{m_X^4} p \cdot q \Big ] .
  \end{split}
\end{equation}
The slashed product can be rewritten into $\slashed{p}\slashed{q} =
p\cdot q-i p^\mu\sigma_{\mu\nu}q^\nu$.  After performing the $p$ integration the
second term would have to be of the form $i q^\mu\sigma_{\mu\nu}q^\nu$ and,
due to the antisymmetry of $\sigma_{\mu\nu}$, such a term will not contribute.
After the operations given above, we obtain
\begin{equation}
  \begin{split}
    \Sigma^P_1(0) &= -\frac{m_{\Delta_i}^2}{2m_X^4}\mz{0}{0}{m_{\Delta_i}}
    - (D-1) \bigg(\frac{1}{2m_X^4}A_0(m_X^2)^2 \\
    & \quad {} + \frac{m_{\Delta_i}^2}{2}\mz{m_X,0}{m_X,0}{m_{\Delta_i}} \\
    & \quad {} - \mz{m_X,0}{m_X}{m_{\Delta_i}}\bigg).
  \end{split} \label{eq:sigma1-initial}
\end{equation}
This may be rewritten in terms of the simple brackets using relations similar
to those in Eq.~(\ref{eq:difference}).

\subsection{Topology 2}
Neglecting the specific form of the vertices, Topology 2 of
\figref{fig:graphs} leads to the second integral in
Eq.~(\ref{eq:central-scalar-loop-integral}).  It can be simplified into
(again making use of the antisymmetry of $\sigma_{\mu\nu}$)
\begin{equation}
  \begin{split}
    \Sigma^P_2(0) &= -\mz{m_X,0}{m_{\Delta_i},0}{m_X}\Big[
      (2-D) p\cdot q \\
    & \quad {} - \frac{2p^2q^2}{m_X^2}
      - \frac{2p^2+q^2}{m_X^2}p\cdot q
      + \frac{p^4q^2}{m_X^4} \\
    & \quad {} + \frac{p^2(q^2+p^2)}{m_X^4} p\cdot q
      + \frac{p^2}{m_X^4} (p\cdot q)^2 \Big] .
  \end{split}
\end{equation}
The result after simplification reads
\begin{equation}
  \begin{split}
    \Sigma^P_2(0) &= \frac{2-D}{2}\mz{0}{m_{\Delta_i},0}{m_X} \\
    & \quad {} + \frac{3-D}{2}\mz{m_X,0}{m_{\Delta_i}}{m_X} \\
    & \quad {} + \frac{m_{\Delta_i}^2}{4m_X^4} \Big (
      2 \mz{m_X}{0}{m_{\Delta_i}} - \mz{m_X}{m_X}{m_{\Delta_i}} \Big ) \\
    & \quad {} + \frac{D-2}{2m_{\Delta_i}^2m_X^2}A_0(m_X^2)A_0(m_{\Delta_i}^2)
      -\frac{1}{4m_X^4} A_0(m_X^2)^2.
  \end{split} \label{eq:sigma2-initial}
\end{equation}

\subsection{Integrals}
For the reader's convenience, we list here the results of the integrals
appearing in the expressions in our convention.  As integrals $A_0(M_A^2)$
appear in the results in the second power, we need to evaluate also the term
linear in $\epsilon$.  This gives
\begin{align}
  A_0(M_A^2)&= Q^{4-D} \int \frac{d^D p}{(2\pi)^D} \frac{1}{p^2-M_A^2}
    \nonumber \\
  &=-i\frac{M_A^2}{(4\pi)^2} \left [ -\frac{1}{\epsilon}+L_A
    - \frac{\epsilon}{2} \left(L_A^2+1+\frac{\pi^2}{6}\right) \right]
    \nonumber \\
  & \quad {} + O\left(\epsilon^2\right),
\end{align}
where
\begin{equation}
L_A = \log\frac{M_A^2}{Q^2} - \log 4\pi + \gamma-1 ,
\end{equation}
with $Q$ being the renormalization scale and $\gamma$ the Euler-Mascheroni
constant.

As was already stated, all of the simple brackets can be obtained from the
double brackets using Eq.~(\ref{D-3}).  Therefore, we give here the result
only for them.  It reads
\begin{equation} \label{eq:double-bracket-result}
  \mz{2M}{M_a}{M_b}=\frac{1}{(4\pi)^4}\left(S(M)-f(a,b)\right)
  + O\left(\epsilon\right) ,
\end{equation}
where
\begin{gather}
  S(M) = -\frac{1}{2\epsilon^2}+\frac{1}{\epsilon} \left( L +
  \frac{1}{2} \right ) - \left ( L^2 + L + \frac{1}{2}
  + \frac{\pi^2}{12} \right ) , \\
  a = \frac{M_a^2}{M^2}, \qquad b = \frac{M_b^2}{M^2} ,
\end{gather}
and the function $f(a, b)$ is given by
\begin{widetext}
  \begin{gather}
    \begin{split}
    f(a,b) = -\frac{1}{2}\log a \log b + \frac{1-a-b}{2\sqrt{q}}
    \Bigg [ &\dilog \left ( -\frac{x_{2}}{y_{1}} \right )
      + \dilog \left( -\frac{y_{2}}{x_{1}} \right )
      - \dilog \left ( -\frac{x_{1}}{y_{2}} \right )
      - \dilog \left ( -\frac{y_{1}}{x_{2}} \right) \\
    & {}+ \dilog \left( \frac{b-a}{x_{2}} \right )
      + \dilog \left ( \frac{a-b}{y_{2}} \right )
      - \dilog \left ( \frac{b-a}{x_{1}} \right )
      - \dilog \left ( \frac{a-b}{y_{1}} \right )
      \Bigg ] ,
    \end{split} \label{eq:fab} \\
    f(b,b) = -\frac{(2 b-1) \left (
      2\dilog \left( \frac{\sqrt{1-4 b}-1}{\sqrt{1 - 4b} + 1} \right )
      + \frac{\pi^2}{6} + \frac{1}{2} \log^2 \left (
      -\frac{\sqrt{1 - 4 b} - 1}{\sqrt{1 - 4 b} + 1} \right ) \right )}
      { \sqrt{1-4b}} - \frac{1}{2} \log ^2(b) . \label{eq:fbb}
  \end{gather}
\end{widetext}
In Eq.~(\ref{eq:fab}) and Eq.~(\ref{eq:fbb}) the quantities $q$, $x_{1,2}$,
and $y_{1,2}$ are defined by
\begin{align}
  q &\equiv 1-2(a+b)+(a-b)^{2} \, , \\
  x_{1,2} &\equiv \frac{1}{2} ( 1 + b- a \pm \sqrt{q}) \, , \\
  y_{1,2} &\equiv \frac{1}{2} ( 1 + a - b \pm \sqrt{q} ) \, .
\end{align}
In addition to Eq.~(\ref{eq:fbb}) giving the value of $f(a,b)$ when
$a = b$, it is helpful to note the additional special cases
\begin{align}
  f(0,0) &= \frac{\pi^2}{6}, \\
  f(0,b) &= \dilog ( 1 - b ) , \\
  f(0 , b^{-1}) &= - \frac{1}{2} \log^2 b - f(0, b) .
\end{align}
\begin{widetext}

\subsection{The kinematic structure of the self-energies}
Rewriting Eq.~(\ref{eq:sigma1-initial}) and Eq.~(\ref{eq:sigma2-initial})
yields the expressions in terms of double brackets,
  \begin{align}
    \begin{split}
      \Sigma^P_1(0) &=
        -\frac{1}{D-3}\frac{m_{\Delta_i}^4}{2m_X^4}\mz{2m_{\Delta_i}}{0}{0}
        - \frac{D-1}{2m_X^4}A_0(m_X^2)^2
        + \frac{D-1}{D-3}(2\mz{2m_X}{m_X}{m_{\Delta_i}}
        - \mz{2m_X}{0}{m_{\Delta_i}}) \\
      & \quad {} + \frac{D-1}{D-3}\frac{m_{\Delta_i}^4}{2m_X^4} (
        2 \mz{2m_{\Delta_i}}{m_X}{0} - \mz{2m_{\Delta_i}}{m_X}{m_X}
        - \mz{2m_{\Delta_i}}{0}{0} )\\
      & \quad {} + \frac{D-1}{D-3}\frac{m_{\Delta_i}^2}{m_X^2} (
        \mz{2m_X}{0}{m_{\Delta_i}} - \mz{2m_X}{m_X}{m_{\Delta_i}}
        + \mz{2m_{\Delta_i}}{m_X}{m_X} - \mz{2m_{\Delta_i}}{m_X}{0} ) ,
    \end{split}\\
    \begin{split}
      \Sigma^P_2(0) &=
        \frac{D-2}{2m_{\Delta_i}^2m_X^2} A_0(m_X^2) A_0(m_{\Delta_i}^2)
        - \frac{1}{4m_X^4} A_0(m_X^2)^2
        + \frac{D-2}{D-3}\frac{m_X^2}{2m_{\Delta_i}^2}(\mz{2m_X}{0}{0}
        - \mz{2m_X}{0}{m_{\Delta_i}} ) \\
      & \quad {} + \frac{m_{\Delta_i}^2}{2m_X^2} (
        \mz{2m_{\Delta_i}}{m_X}{0} - \mz{2m_{\Delta_i}}{m_X}{m_X} )
        + \frac{1}{D-3}\frac{m_{\Delta_i}^2}{2m_X^2}(
        \mz{2m_X}{0}{m_{\Delta_i}} - \mz{2m_X}{m_X}{m_{\Delta_i}} ) \\
      & \quad {} - \frac{D-2}{2(D-3)}\mz{2m_{\Delta_i}}{m_X}{0}
        - \frac{1}{2}(2\mz{2m_X}{m_X}{m_{\Delta_i}}
        - \mz{2m_X}{0}{m_{\Delta_i}} ) \\
      & \quad {} + \frac{1}{D-3}\frac{m_{\Delta_i}^4}{4m_X^4}  (
        2 \mz{2m_{\Delta_i}}{m_X}{0}-\mz{2m_{\Delta_i}}{m_X}{m_X}) .
    \end{split}
  \end{align}
  Using the explicit expression for the double brackets,
  Eq.~(\ref{eq:double-bracket-result}), $\Sigma_1^P(0)$ and $\Sigma_2^P(0)$
  are then finally found to be given by (where $s_i =
  \frac{m_{\Delta_i}^2}{m_X^2}$ as above)
\begin{align}
  \begin{split}
    (4\pi)^4\Sigma(0)^P_1 &= -\frac{3}{2\epsilon} + 3L_X - 2
      + \frac{s_i^2}{2} \left [ \frac{1}{2\epsilon^2}
      - \frac{1}{\epsilon} \left ( L_{\Delta_i} - \frac{1}{2} \right )
      + \left ( L_{\Delta_i}^2 - L_{\Delta_i} + \frac{3}{2} + \frac{\pi^2}{12}
      \right ) \right] \\
    & \quad {} + 3 \big ( f(0,s_i) - 2 f(1,s_i) \big )
      + \frac{3}{2} s_i^2 [ f(s_i^{-1},s_i^{-1})
      - 2f(0,s_i^{-1})  ] \\
    & \quad {} + 3s_i [ f(1,s_i) - f(0,s)
      - f(s_i^{-1},s_i^{-1})+f(0,s_i^{-1})]
      +2s_i^2f(0,0),
  \end{split} \\
  \begin{split}
    (4\pi)^4\Sigma(0)^P_2 &= \frac{3}{4\epsilon}
      - \frac{1}{2} \left [ L_X + 2 L_{\Delta_i} - (L_{\Delta_i}-L_X)^2 - 1
      \right ]
      - \frac{s_i^2}{4} \left [ \frac{1}{2\epsilon^2}
      - \frac{1}{\epsilon} \left ( L_{\Delta_i} - \frac{1}{2} \right )
      + \left ( L_{\Delta_i}^2 - L_{\Delta_i} + \frac{3}{2}
      +\frac{\pi^2}{12} \right ) \right ]\\
    & \quad 
      - s_i^{-1} [f(0,0) - f(0,s_i)] + f(0,s_i^{-1}) + f(1,s_i) - \frac{1}{2} f(0,s_i)\\
    & \quad {} - \frac{s_i}{2}[f(s_i^{-1},0)
      - f(s_i^{-1},s_i^{-1}) + f(0,s_i) - f(1,s_i) ]- \frac{s_i^2}{4} [ 2f(0,s_i^{-1})
      - f(s_i^{-1},s_i^{-1})].
  \end{split}
\end{align}
\end{widetext}
Note that the individual diagrams are UV divergent, with the divergent
terms given by Eq.~(\ref{eq:central-scalar-divergences}) and
Eq.~(\ref{eq:edge-scalar-divergences}).  However, as noted in
\secref{sec:calculation}, their combination appearing in
Eq.~(\ref{eq:integral-sum-definition}) yielding the total contribution
to the RH neutrino mass matrix is finite and compact,
\begin{align} \label{eq:I3intermsofs}
    I_3(s) &= 1 + 2 \log s + s ( 1 - 2 s ) \log^2 s \\
    & \quad {} + 2\left ( s^{-1} - 1 \right ) \Big [ f(0,0) ( 1 + s + s^2 )+ 2 s f(1,s)
      \nn\\
    & \quad {}  + f(0,s) ( 1 + s ) ( 1 + 2 s) +s^2 f(s^{-1}, s^{-1}) \Big ]\nn .
\end{align}

\bibliographystyle{h-physrev5}
\bibliography{bibliography-witten}

\end{document}